\documentclass[aps, twocolumn,superscriptaddress]{revtex4-2} \usepackage{natbib}
\usepackage{import}
\usepackage{xifthen}
\usepackage{transparent}
\usepackage[colorlinks=true]{hyperref}
\usepackage{array}
\usepackage{tikz}
\usepackage{verbatim, bm}
\usepackage{physics}
\usepackage{textcomp}
\usepackage{nicefrac}
\usepackage{mathrsfs}  
\usepackage[normalem]{ulem}
\usepackage{mathtools}
\usepackage{amsfonts}
\usepackage{amssymb}
\usepackage{color}
\usepackage{amsthm}
\usepackage{booktabs}
\usepackage{slashed}
\usepackage{amsmath}
\usepackage{wrapfig}
\usepackage{dsfont}
\usepackage{esint}
\usepackage{siunitx}

\hypersetup{
    unicode=false,          
    pdftoolbar=true,        
    pdfmenubar=true,        
    pdffitwindow=false,     
    pdfstartview={FitH},    
    pdfsubject={},          
    pdfcreator={},          
    pdfproducer={},         
    pdfkeywords={} {} {},   
    pdfnewwindow=true,      
    colorlinks=true,        
    linkcolor=magenta,      
    citecolor=blue,         
    filecolor=magenta,      
    urlcolor=blue           
} 
\allowdisplaybreaks

\newcommand{\bq}{{\bf q}}

\newcommand{\bz}{{\bf z}}

\newcommand{\ba}{{\bf a}}

\newcommand{\bp}{{\bf p}}
\newcommand{\br}{{\bf r}}
\newcommand{\bR}{{\bf R}}
\newcommand{\bs}{{\bf s}}

\renewcommand{\bar}{\overline}

\renewcommand{\hat}{\widehat}
\renewcommand{\leq}{\leqslant}
\renewcommand{\geq}{\geqslant}
\renewcommand{\Re}{\operatorname{Re}}

\newcommand{\UU}{\operatorname{U}}

\newcommand{\sgn}{\operatorname{sgn}}

\usepackage{orcidlink}

\begin{document}
\title{Anyonic Neural Network Wavefunctions}

\author{Andres Perez Fadon \orcidlink{0009-0005-7615-879X}}
\affiliation{Department of Physics, Imperial College London, London SW7 2AZ, United Kingdom}
\email{andres.perez-fadon19@imperial.ac.uk}

\author{Pok Man Tam \orcidlink{0000-0001-9620-4689}}
\affiliation{Princeton Center for Theoretical Science, Princeton University, Princeton, NJ 08544, USA}
\affiliation{Leinweber Institute for Theoretical Physics, University of Chicago, Chicago, IL 60637, USA}
\email{pokman.tam@uwaterloo.ca}

\author{Leyna Shackleton \orcidlink{0000-0002-1971-6426}}
\affiliation{Department of Physics, Massachusetts Institute of Technology, Cambridge MA 02139, USA}
\email{lshackle@mit.edu}

\begin{abstract}
The macroscopic behavior of a non-interacting gas of Abelian anyons has remained unsettled for many decades. Prior numerical methods employing exact diagonalization in the continuum and on the plane have been limited to small system sizes, at most $N = 4$ anyons. Using the rapidly developing technology of neural network wavefunctions, we develop variational ansatze appropriate for studying continuum models of the Abelian anyon gas. The neural network architecture, which we call ``AnyNet'', is capable of exactly enforcing the non-analytic short-distance behavior induced by the anyonic statistics. We demonstrate excellent agreement with exact diagonalization for $N=4$ anyons, obtain accurate results up to $N = 16$ for intermediate statistics, and study up to $N=20$ for almost-fermionic anyons, matching perturbative results within a 0.4\% discrepancy.
We use this methodology to study anyons in a harmonic trap, and anyonic atoms as a model for anyon binding to impurities. 
We study the structure of anyonic wavefunctions as the statistics parameter is varied.
We focus on the single-particle density, the pair density, the angular momentum, and the scaling of the wavefunction as multiple anyons coalesce.
This work opens doors to understanding the macroscopic behavior of anyons, for which rich phases have been predicted theoretically.
\end{abstract}
\maketitle

\section{Introduction}
Two-dimensional systems admit particles, called \textit{anyons}, whose exchange statistics are beyond the conventional paradigms of bosons and fermions~\cite{leinaas1977, wilczek1982, wu1984a, wu1984}. 
Anyons arise most prominently in the fractional quantum Hall effect (FQHE)~\cite{Tsui_Stormer_FQH_1982, laughlin1983}, where partially-filled Landau levels give rise to incompressible phases of matter with excitations possessing anyonic statistics~\cite{halperin1984,arovas1984, wilczek1982magnetic}. Recently, such exotic statistics effects have been directly observed in fractional quantum Hall interferometry as well as current noise experiments ~\cite{Nakamura2020Braiding,Bartolomei2020Collisions,
Nakamura2023TwoFifths,
Kim2024BilayerInterferometry,
Werkmeister2025TelegraphBraiding,
Ghosh2025ChiralBraiding}.
While the strong magnetic fields necessary for the FQHE generally quench the kinetic energy of the emergent anyons, there is growing interest in the many-body physics of anyons due to the discovery of the fractional quantum anomalous Hall effect~\cite{cai2023alt,park2023,Li2023_QAH,zeng2023alt, lu2024a, xu2026signaturesunconventionalsuperconductivitynear}, where the absence of a magnetic field and the presence of a lattice allow the anyons to acquire a dispersion \cite{Miguel_anyondispersion_2026, Khalaf_dispersion_2026a, Khalaf_dispersion_2026b, iyer2026dispersionanyonblochbands}.
Many recent theoretical proposals have been made regarding the macroscopic behavior of a finite density of anyons ~\cite{shi2024, shi2025, shi2025a,divic2025a, mehta2026, shi2026, shi2026a, senthil2026, han2026a, Wang_Zhang_ASF_2026, zhang2026, zhang_holon_2026, lotric2026, fan2026hiddenweakpairingsuperconductivitynoninteracting, Pichler2026MicroscopicMechanism,wang2026topologicalsuperconductivityabelianfractional}. 
Despite extensive analysis, the fate of a ``free'' anyon gas remains unsettled.

In contrast to bosons and fermions, the ground state of a system of non-interacting anyons cannot be simply constructed from single-anyon eigenstates. This can be easily seen via a singular gauge transformation that translates the multivaluedness of the anyon wavefunction to a long-ranged interaction generated by a $\UU(1)$ Chern-Simons gauge field. 
This makes the physics of ``free'' anyons, where the only interaction is the statistical gauge interaction, non-trivial. As a result, the macroscopic behavior of a non-interacting finite density of anyons remains unsettled. Early theoretical studies predicted that a finite density of anyons will give rise to a superfluid state~\cite{laughlin1988, fetter1991, hanna1989, chen1989}, although these calculations are uncontrolled and assume small fluctuations around a mean-field state. Numerical studies of these systems~\cite{kallin1993, feder1995, xie1990} give results consistent with superfluidity, but are restricted to small systems.

Numerical studies of anyons fall largely into two categories.
On the one hand, the well-established platform of the FQHE and Landau level
physics allows one to study anyons as quasiparticle excitations in strongly
correlated electronic systems. This includes numerical studies of trial
wavefunctions~\cite{TserkovnyakSimon2003,BarabanEtAl2009,GattuJain2025MolecularAnyons,LiEtAl2026Geometric},
exact diagonalization (ED)
studies~\cite{ProdanHaldane2009,Wojs2010Skyrmions,StorniMorf2011,Moller2011NeutralFermion,Kapit2012NonAbelianBraiding,Cooper2015ExclusionStatistics,Liu2015QuasiholesFCI,Jaworowski2019QuasiholesC2,XuEtAl2025AnyonClusters},
as well as tensor
network~\cite{ZaletelMong2012,ZaletelMongPollmann2013,WuEtAl2014,WuEtAl2015,GrushinEtAl2015,KjallEtAl2018,MacalusoEtAl2020,WangZaletel2026}
and neural-network variational Monte Carlo
methods~\cite{qian2025, NazaryanEtAl2025, teng2025, gattu2025, luo2025tmote2, li2025deep, zhang2025transformer, perez2026extracting, abouelkomsan2026topological, abouelkomsan2026crystallization,GrahamEtAl2026}.

Studying anyons in the FQHE, however, gives one limited control over the effective anyon Hamiltonian and interactions. Moreover, simulating emergent anyons in microscopic electronic systems severely constrains the number of anyons that can be studied. Tackling the problem directly with an intrinsically anyonic Hamiltonian is therefore of interest. Numerical studies of these models have, however, been limited and underexplored over the past several decades. Early ED works on continuum models of anyons were able to access $N=3$ and $N=4$ anyons in a harmonic trap~\cite{sporre1991, sporre1992, murthy1992}, and up to $N=8$ on a sphere~\cite{xie1990}. Lattice models of anyons were also able to access up to $N=10$ via ED~\cite{kallin1993, feder1995}. Modern numerical methods have yet to be applied to these anyonic models.
We note that neural-network states with anyonic symmetries have been constructed on lattices~\cite{vieijra2020, luo2021gauge}. 
However, the anyonic structure enters through fusion-category constraints on lattice degrees of freedom, rather than through particles with fractional exchange statistics as considered here.

Among the modern methods available to us, neural-network variational Monte Carlo (NNVMC) is one of the most promising~\cite{carleo2017}. 
NNVMC leverages the expressivity of neural networks to parameterize a large class of variational states, which can be optimized against the expectation value of a Hamiltonian to obtain an approximation to the ground state. 
While NNVMC has been applied both to lattice~\cite{carleo2017, nomura2017, luo2019a, choo2019, roth2023, roth_large_scale_2026} and continuum systems, including molecules~\cite{hermann2020, pfau2020ab, glehn2023, gao2024pfaffians, pfau2024excited, carr2026neural}, solids and electron gases~\cite{li2022deepsolid, cassella2023, lou2024neural, pescia2024, smith2024, lou2026uncovering}, as well as particles in harmonic traps~\cite{ruggeri2018, saito2018, yang2020, adams2021, keeble2023, kim2024, perez2025interaction, zaklama2026large}, 
the most dramatic gains have arguably been for the continuum systems, as alternative methods for them are more limited. Combined with the strongly correlated nature of the free anyon gas, which limits the applicability of other variational methods, neural-network ansatze are well-positioned to make substantial contributions to our understanding of anyonic systems.

In this work, we present the first application of NNVMC to fundamental anyonic systems. We use a singular gauge transformation to convert the system to a model of fermions interacting via a statistical gauge interaction, although we stress that all aspects of our computation are mathematically identical to working in a gauge with a multivalued wavefunction. 
Conventional fermionic neural-network ansatze, such as FermiNet \cite{pfau2020ab, spencer2020better, glehn2023}, perform poorly when directly applied to this model due to the singular short-distance behavior of the statistical gauge interaction. 
As we will discuss, it is insufficient to impose the non-analytic behavior with a Jastrow factor over pairs, in contrast to previously studied fermionic systems.
One of the primary contributions of this work is developing an ansatz that retains the expressivity of modern neural-network ansatze while exactly enforcing the desired short-distance behavior.

\subsection{Overview}
\begin{figure}
    \centering
    \includegraphics[width=\linewidth]{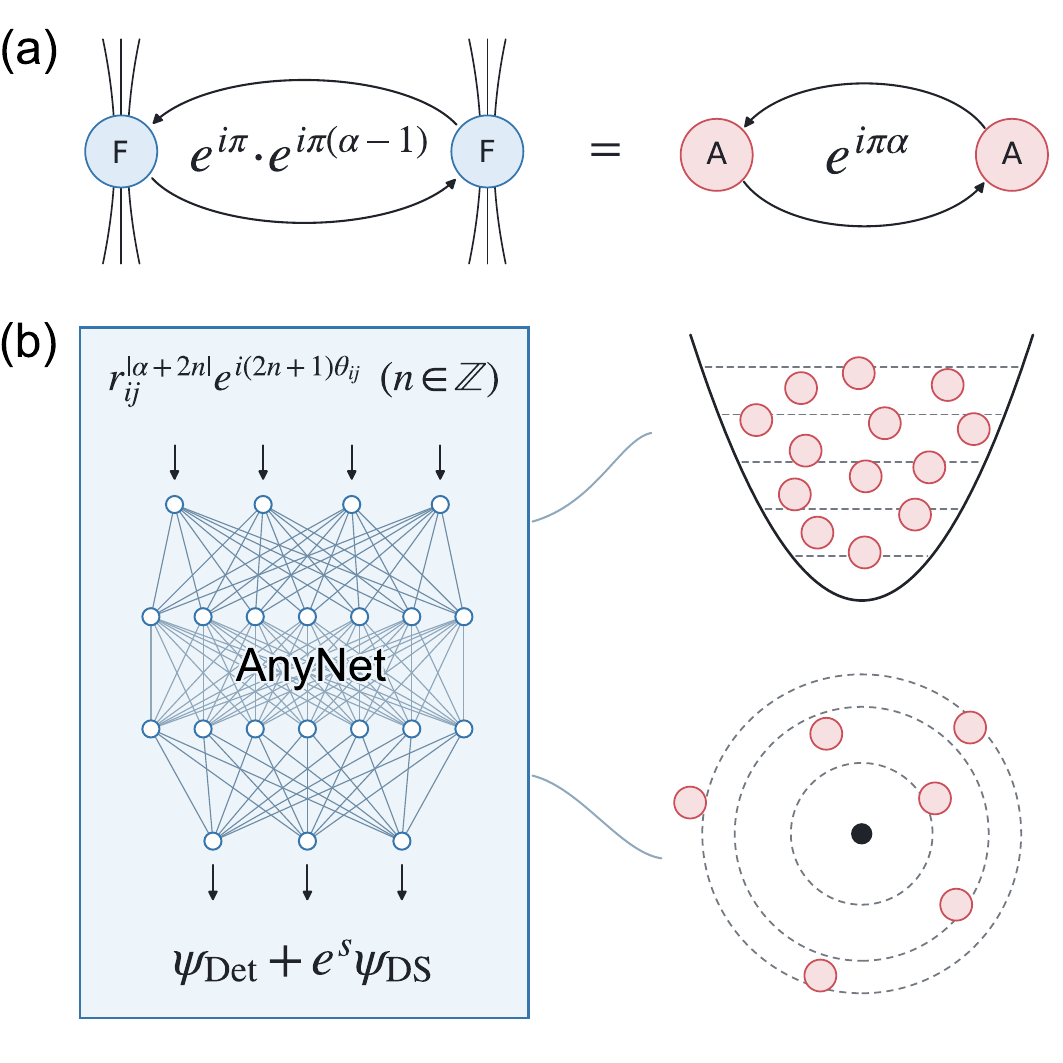}
    \caption{Schematic overview of this work, where a neural network architecture termed ``AnyNet" is proposed to investigate generic many-anyon systems. (a) Essential to our methodology is the statistical transmutation, where anyons with exchange phase $e^{i\pi\alpha}$ are identified as flux-binding fermions, and hence an anyonic system is converted into an interacting fermionic system with long-range couplings to a statistical gauge field. (b) Depending on the statistics parameter $\alpha$, anyons have distinctive short-distance coalescence behavior, which have been engineered into a modified FermiNet-type neural network architecture ($\psi_{\rm Det}$). This is further combined with a Deep Set architecture $(\psi_{\rm DS})$ that allows for an efficient representation of an almost-bosonic system. Altogether, the AnyNet ansatz is demonstrated to be effective and expressive for simulating the classic problem of the anyon gas in a harmonic trap, as well as a model of anyonic atoms, for a wide range of anyon number, statistics parameters, and in the presence of interactions.}
    \label{fig:overview}
\end{figure}

Our paper is structured as follows. In Section~\ref{chap:setup}, we define our anyonic systems of interest and review what is known analytically about the structure of the ground states. 
In Section \ref{chap:nn_general}, we provide an overview of neural-network ansatze, including the FermiNet for conventional correlated electron problems, and a discussion of further modifications necessary for studying anyons. This leads us to propose our neural-network ansatz, termed ``AnyNet", whose architecture is detailed in Section \ref{chap:architecture}. In this work, we benchmark and apply this ansatz to a number of different systems. Major results are presented in Section~\ref{sec: NNVMC results}.
In Section \ref{chap:ED_comparison} we first study $N=4$ anyons in a harmonic trap, and compare our results to ED, following the method in Ref.~\cite{sporre1992}, finding excellent agreement. We then extend to larger $N$ values that have not been studied before from first-principles. Particularly, for $N=10$, we provide a detailed study of level crossings and anyon density profiles (Section~\ref{chap:observables}). For statistics near the fermionic limit, 
we analytically obtain perturbative results
and find excellent agreement with our numerical AnyNet results up to $N=20$, see Section \ref{sec: fermionic perturbation theory}.
In Section \ref{chap:coalescence_results} we study the scaling of the neural-network wavefunction at the coalescence of multiple anyons, and compare with non-relativistic conformal field theory predictions.
Lastly, in Section \ref{chap:anyonic_atoms} we study both non-interacting and interacting anyons in a Coulomb trap, as a demonstration that our method generalizes to other anyonic systems.
We conclude in Section \ref{chap:conclusion} with future directions that become possible extending the ideas presented in this study.

\section{Hamiltonian and Setup}
\label{chap:setup}
Here we provide the theoretical background for studying many-anyon systems. In particular, we explain the statistical transmutation approach that enables us to transform an anyonic problem into an interacting fermionic problem, which is later on solved with NNVMC. We also specify here the two model Hamiltonians to be investigated using NNVMC. 
For anyons in a harmonic trap, we review some known theoretical results that provide important guidance for the construction of our neural network ansatze.

\subsection{Statistical transmutation}
Consider $N$ identical, non-relativistic, Abelian anyons in two spatial dimensions. 
As a defining feature of Abelian anyons, their physical wavefunction $\psi(\{\mathbf r_i\})$ is multivalued, such that a counterclockwise exchange (assuming the exchange loop does not enclose any other anyons) produces
\begin{equation}
\begin{split}
    \psi_{A}(\ldots,\mathbf r_i,\mathbf r_j,\ldots)
\longrightarrow
&\;\psi_{A}(\ldots,\mathbf r_j,\mathbf r_i,\ldots)\\
=&\;e^{i\pi\alpha}
\psi_{A}(\ldots,\mathbf r_i,\mathbf r_j,\ldots),
\end{split}
\label{eq:anyon-exchange}
\end{equation}
where $0\leq\alpha\leq 1$ is the exchange statistics parameter.
The subscript $A$ denotes that the state is in the so-called ``anyon gauge".
The Hamiltonian acting on the multivalued anyon wavefunction takes the following form in general
\begin{equation}
H_{A}
=
\sum_{i=1}^N
\left[
-\frac{\hbar^2}{2m}\nabla_i^2+ V(\br_i)
\right]+U(\{\br_i\}).
\label{eq:anyon-gauge-hamiltonian}
\end{equation}
where $m$ is the anyon mass, $V(\br_i)$ the external one-anyon potential and $U(\{\br_i\})$ describes interaction between anyons.

For our purpose of implementing NNVMC, it is useful to trade the multi-valuedness of wavefunction for a coupling to a statistical gauge field. Writing $z_i=x_i+i y_i$ and $\theta_{ij}=\arg(z_i-z_j)$, we define a single-valued, antisymmetric wavefunction $\psi_F$ through a gauge transformation:
\begin{equation}
\psi_{A}(\{\br_i\})=e^{i(\alpha-1)\sum_{i<j}\theta_{ij}}\psi_{F}(\{\br_i\}).
\label{eq:singular-gauge-transformation}
\end{equation}
This is known as statistical transmutation, which is pictorially represented in Fig.~\ref{fig:overview}(a) as flux-attachment.
The phase factor in Eq.~\eqref{eq:singular-gauge-transformation} combines with the fermionic exchange sign from $\psi_F$ to produce the anyonic exchange phase of $\psi_{A}$. 
We say we are adopting the ``fermion gauge" when working with the single-valued fermionic wavefunction $\psi_F$, and quantities in this gauge will carry a subscript $F$.
Under this gauge transformation, the anyonic  Hamiltonian transformed from Eq.~\eqref{eq:anyon-gauge-hamiltonian} to
\begin{equation}
H_{F}=
\sum_{i=1}^N\left[ \frac{\hbar^2}{2m}
\left(-i\nabla_i+\mathbf a_i\right)^2 +V(\br_i)\right]
+U(\{\br_i\})
\label{eq:anyon-hamiltonian}
\end{equation}
where
\begin{equation}
\mathbf a_i
=
(\alpha-1)\sum_{j\neq i}
\frac{\hat{\mathbf z}\times(\mathbf r_i-\mathbf r_j)}
{|\mathbf r_i-\mathbf r_j|^2}.
\label{eq:statistical-gauge-field}
\end{equation}
This is describing an interacting fermionic problem where each fermion is coupled to a gauge field $\ba_i$ that produces the effect of statistical transmutation between anyons and fermions. The statistical interaction is singular, long-ranged, and contains a 3-body interaction (from the ${\mathbf a^2_i}$ term). The operator is defined away from coincident particles, while its physical content at coincidence is fixed by a choice of self-adjoint extension. 
This gauge transformation demonstrates that, while one may have expected the difficulty in designing a variational anyonic wavefunction to be in accommodating its multi-valued nature, such a condition can easily be imposed on a wavefunction via the gauge transformations described above. The true complexity lies in the fact that, even for a ``free" (i.e. $V=U=0$) anyonic Hamiltonian in Eq.~\eqref{eq:anyon-gauge-hamiltonian} with a multi-valued wavefunction (or working in the ``anyon gauge"), the problem is equivalent to solving a strongly-correlated system of fermionic particles in the ``fermion gauge". The goal of this work is to introduce an effective variational ansatze for $\psi_F$, aided by the recent development of neural network wavefunctions, to simulate a large family of many-anyon systems following the above recipe of statistical transmutation. We apply NNVMC specifically to two anyonic systems: the free-anyon gas in a harmonic trap and the anyonic atom.

\subsection{Harmonic trap: theory and previous works}\label{sec: theory and previous works}

For the first model, we consider a harmonic trap potential $V(\br) = \frac{1}{2}m\omega^2|\br|^2$, and no interaction between anyons ($U=0$). For notational simplicity, moving forward we set $\hbar=m=\omega=1$.
We choose to study the harmonic trap problem in part because the $N$-anyon ground state energy is of a fundamental importance related to the generic short-distance behavior of the wavefunction as $N$ anyons coalesce, as shown in Section \ref{chap:coalesecence}.

\subsubsection{Exact eigenstates}
Let us first review the analytically known exact eigenstates for the harmonic trap problem. For our consideration here, it is most convenient to adopt the ``anyon gauge" (see Eq.~\eqref{eq:anyon-gauge-hamiltonian}), where the anyonic statistics is encoded in the boundary condition of the wavefunction as in Eq.~\eqref{eq:anyon-exchange}.

For $N=2$ anyons, the entire spectrum is exactly known \cite{wilczek1982, leinaas1977, wu1984}. Introducing the center-of-mass (CM) coordinate $\bR=\frac{1}{2}(\br_1+\br_2)$ and the relative coordinate $\br=\br_1-\br_2$, we have $H_A = H_{\rm CM}+H_{\rm rel}$, with $H_{\rm CM} = -\frac{1}{4}\nabla_\bR^2 + \abs{\bR}^2$ and $H_{\rm rel} = -\nabla_\br^2+\frac{1}{4}\abs{\br}^2$. Solving $H_{\rm CM}$ is straightforward and gives the spectrum $E_{\rm CM}^{(m,p)} = 2m+\abs{p}+1$, with $m\in\mathbb{Z}^+_0$ the radial quantum number and $p\in\mathbb{Z}$ the angular momentum in the CM sector. As for the relative sector, expressing $\Psi_{\rm rel}(r,\theta) = \mathcal{R}(r) e^{i\ell\theta}$, the anyonic statistics in Eq.~\eqref{eq:anyon-exchange} requires the relative angular momentum $\ell = 2q+\alpha$ for $q\in\mathbb{Z}$. The Schr\"{o}dinger equation becomes
\begin{equation}\label{eq: radial Schrodinger equation}
    \frac{\partial^2 \mathcal{R}}{\partial r^2} + \frac{1}{r} \frac{\partial \mathcal{R}}{\partial r} +\left(E_{\rm rel} - \frac{\ell^2}{r^2} - \frac{1}{4}r^2 \right) \mathcal{R} = 0.
\end{equation}
Its solution takes the form $\mathcal{R}(r) = r^{\abs{\ell}}e^{-r^2/4}L^{(\abs{\ell})}_n(\frac{r^2}{2})$, where the first two factors are obtained from the asymptotic behaviors as $r\rightarrow 0$ and $r\rightarrow\infty$, and $L^{(\abs{\ell})}_n(x)$ is the generalized Laguerre's polynomial whose degree $n=\frac{1}{2}(E_{\rm rel}-\abs{\ell}-1) \in\mathbb{Z}^+_0$ is a non-negative integer to ensure regularity of the solution. It is important to remark that we have chosen to focus on ``hard-core anyons", which requires the wavefunction to vanish as anyons coalesce. Therefore, while the $r\rightarrow0$ asymptotic behavior of Eq.~\eqref{eq: radial Schrodinger equation} generally allows for $\mathcal{R} \sim r^{\pm\abs{\ell}}$ (both satisfy $L^2$ integrability as long as $\abs{\ell}<1$), the hard-core boundary condition selects $\mathcal{R} \sim r^{\abs{\ell}}$. This is also consistent with the short-distance behavior of the two-anyon problem with no external potential as analyzed in Ref. \cite{arovas1985} (see also Appendix \ref{chap:short_range}). Altogether, we obtain the full $N=2$ spectrum as
\begin{equation}\label{eq: 2-anyon exact}
    E_2^{(m,n,p,q)} = 2(m+n+1) +\abs{p}+\abs{2q+\alpha}.
\end{equation}
The entire spectrum thus depends linearly on the statistics parameter, which is a unique feature for $N=2$. For $\alpha\in[0,1]$, we would like to highlight two low-energy states with
\begin{equation}\label{eq: two energies for 2 anyons}
    E_2^{(0,0,0,0)} = 2+\alpha,\quad\text{and}\quad  E_2^{(0,0,0,-1)} = 4-\alpha.
\end{equation}
Notice that their wavefunctions vanish as $\rho^{\alpha}$ and $\rho^{2-\alpha}$, respectively, as the distance $\rho$ between the two anyons vanishes.

For $N>2$, while the full spectrum can no longer be solved analytically, subsets of eigenstates are analytically known, with energy varying linearly with the statistics parameter \cite{wu1984, polychronakos1991exact, murthy_exact1992}. Below we focus on the simplest class that includes the multi-anyon ground state near the bosonic limit. 
Rewriting $H_A$ in Eq.~\eqref{eq:anyon-gauge-hamiltonian} in terms of complex coordinates $z_i = x_i+iy_i$ and $\bar{z}_i = x_i-i y_i$, it is straightforward to check that 
\begin{equation}\label{eq: holomorphic states}
\Psi(\{z_i,\bar z_i\})
= P(\{z_i\})
e^{-\frac{1}{2}\sum_i \abs{z_i}^2}\prod_{i<j}(z_i-z_j)^\alpha
\end{equation}
is always an eigenstate as long as $P(\{z_i\})$ is a holomorphic, symmetric and homogeneous polynomial. Denoting the degree of $P(\{z_i\})$ as $d$, the corresponding energy is
\begin{equation}\label{eq: holomorphic energy}
    E^{(d)}_{N} = N+\binom{N}{2}\alpha+d.
\end{equation}
In the rest of the work, we refer to such states (particularly for $d=0$, $P(\{z_i\})=1$) as the ``holomorphic states". 
For any finite $N$, with $\alpha$ smaller than some finite threshold $\alpha_*(N)$, we observe that the holomorphic state with energy $E^{(0)}_N$ is indeed the ground state for $N$ anyons, with $\alpha_*(N)$ a decreasing function.  
But for $\alpha>\alpha_*$, the ground state becomes more complicated and its energy varies nonlinearly with $\alpha$. 
For $N=3$ and 4, previous numerical works have found this crossing (from the linear holomorphic branch to the nonlinear branch) to happen around $\alpha_* \approx 0.7$ and 0.55, respectively \cite{sporre1991, sporre1992, sporre1993,  mashkevich1995}. 
In this work, we not only verify these results using NNVMC, but further obtain the nonlinear ground states (over the entire range of $\alpha$) for a large range of anyon number $N$.

In some literature \cite{mashkevich1995, myrheim2002anyons,dft_anyon_Jain2021}, the states in Eq.~\eqref{eq: holomorphic states} are also referred to as the type-I states. Similarly, there are ``anti-holomorphic states", or type-II states, which take the form
\begin{equation}
\Psi(\{z_i,\bar z_i\})
= P(\{\bar{z}_i\})
e^{-\frac{1}{2}\sum_i \abs{z_i}^2}\prod_{i<j}(\bar{z}_i-\bar{z}_j)^{2-\alpha}
\end{equation}
and obey the same statistics as in Eq.~\eqref{eq:anyon-exchange}. This can be understood as the time-reversed partner of the above holomorphic state but with the replacement $\alpha \rightarrow 2-\alpha$ to maintain the same statistics.
However, type-II states have higher energy than type-I states.

Apart from exact eigenstates, existing works have also studied the anyon gas perturbatively near the fermionic or the bosonic limit \cite{chou1991, chitra1992, nishida2007}, semiclassically in the large-$N$ limit using density functional theory \cite{chitra1992, correggi2017local,correggi2019,dft_anyon_Jain2021, girardot2021semiclassical, levitt2026, komargodski2026anyoncrystallizationstatistics}, and by constructions of trial states to bound the ground-state energy \cite{lundholm2017}. For an overview of existing works on anyon gas, we refer interested readers to the review article in Ref. \cite{Lundholm_review}.

\subsubsection{Connection to conformal field theory}
\label{chap:cft}
While in the above we have focused on identifying subsets of exact eigenstates, there is a separate stream of work that casts the problem of non-interacting anyons in a harmonic trap in a broader context in connection to non-relativistic conformal field theory (NRCFT). 
As pointed out by Nishida and Son in Ref.~\cite{nishida2007}, and further discussed by others in Refs.~\cite{doroud2016superconformal, doroud2018conformal}, the energy eigenvalues for the anyon-in-harmonic-trap problem are equivalent to scaling dimensions of primary operators in NRCFTs. Below we briefly review this perspective, which sheds light on both the structure of the energy spectrum and the short-distance anyonic coalescence behaviors. 

Let us adopt a second-quantized formulation for the time being. A complementary perspective in a first-quantized formulation is provided in Appendix~\ref{chap:NRCFT_app}. 
In the fermionic gauge, the free-anyon Hamiltonian without an external potential reads
\begin{equation}\label{eq: second quantized H}
    H_0 = \frac{1}{2}\int d^2\br (\mathcal{D}_i\psi)^\dagger(\mathcal{D}_i\psi),\;\mathcal{D}_i\psi = (\partial_i+ia_i)\psi,
\end{equation}
where $a_i(\br) = (\alpha-1)\int d^2\br'\frac{[\hat{\bz}\times(\br-\br')]_i}{\abs{\br-\br'}^2}n(\br')$ is the statistical gauge potential coupled to the fermionic field $\psi$, and $n(\br)=\psi^\dagger(\br)\psi(\br)$ is the number density operator. Let us further introduce
\begin{equation}\label{eq: C and D operators}
    C = \frac{1}{2}\int d^2\br\; \abs{\br}^2n(\br),\;D = \int d^2\br\; r_i j_i(\br),
\end{equation}
with $j_i = -\frac{i}{2}(\psi^\dagger\mathcal{D}_i\psi - (\mathcal{D}_i\psi)^\dagger \psi)$ the gauge-invariant current operator. The operator $D$ is the generator of dilatation, with $e^{-i\lambda D}\psi(t,\br)e^{i\lambda D} = e^\lambda\psi(e^{2\lambda}t, e^\lambda\br)$, and $C$ is the generator of special conformal transformations \cite{nishida2007}. 
Combining Eq.~\eqref{eq: second quantized H} and Eq.~\eqref{eq: C and D operators}, we can see that they form a closed algebra with $H_0$,
\begin{equation}\label{eq: conformal algebra}
    [H_0,C]=-iD,\;\;[C,D]=2iC,\;\;[D,H_0]=2iH_0,
\end{equation}
which is part of the Schr\"{o}dinger algebra that characterizes a non-relativistic conformal symmetry. Notice that $C$ is also the operator that corresponds to the harmonic potential. 

In a NRCFT, a primary operator $\mathcal{O}(t, \br)$ satisfies 
\begin{equation}\label{eq: primary operator}
    [C,\mathcal{O}]=0, \;\;[D,\mathcal{O}]=i\Delta_{\mathcal{O}}\mathcal{O},
\end{equation}
where $\mathcal{O}\equiv \mathcal{O}(0,0)$ is the short-hand notation for the operator evaluated at the spacetime origin, and $\Delta_\mathcal{O}$ is the corresponding scaling dimension. For any $N$-particle primary operator $\mathcal{O}^\dagger$, Nishida and Son constructed the following state that can be identified as an $N$-anyon energy eigenstate for the harmonic trap problem \cite{nishida2007}:
\begin{equation}\label{eq:nrcft-state}
    \ket{\Psi_\mathcal{O}} = e^{-H_0}\mathcal{O}^\dagger\ket{0},\;\;(H_0+C) \ket{\Psi_\mathcal{O}} = \Delta_\mathcal{O}\ket{\Psi_\mathcal{O}}.
\end{equation}
This can be easily verified using Eq.~\eqref{eq: conformal algebra} and Eq.~\eqref{eq: primary operator}, together with the Baker–Campbell–Hausdorff formula, and $H_0\ket{0}=C\ket{0}=D\ket{0}=0$ for the vacuum $\ket{0}$.

From this operator-state correspondence, the task of obtaining the energy of anyons in a harmonic trap is thus mapped to the problem of calculating the scaling dimension of some $N$-particle primary operators in the theory of Eq.~\eqref{eq: second quantized H}. Particularly, for free fermions in a harmonic trap ($\alpha=1$), with $N=2$ there are two primary operators: $\mathcal{O}^\dagger_{\pm}=\psi^\dagger(\partial_x\pm i \partial_y)\psi^\dagger$ with scaling dimension $3$ (recall $\Delta_\psi = 1$), which corresponds to the obvious fact that with two fermions in a harmonic trap there are two degenerate ground states with energy $3$ (in units of $\hbar\omega$). For the anyonic problem, one can do diagrammatic perturbation calculations to obtain the quantum correction (or anomalous scaling dimension) due to the coupling to the statistical gauge field in Eq.~\eqref{eq: second quantized H}. For $N=2$, as detailed in Ref. \cite{nishida2007}, one obtains $\Delta_{\mathcal{O}_{\pm}} = 3\mp(\alpha-1)$, consistent with the exact result in Eq.~\eqref{eq: 2-anyon exact}. In principle, such diagrammatic calculations can be generalized to arbitrary $N$ and allow for a perturbative calculation for the $N$-anyon energy. This was done explicitly for $N=2,3,4$ in Ref.~\cite{nishida2007}. In Sec. \ref{sec: fermionic perturbation theory}, we present an elementary approach to obtain the perturbative energy near the fermionic limit for generic $N$,  and use that to benchmark our NNVMC results for $N$ up to 20.

Notice that while the above discussion is formulated as fermions coupled to the statistical gauge field, there is an equivalent formulation in terms of bosons, which is suited for doing perturbative analysis for the anyon gas near the bosonic limit, supporting the holomorphic state (see Eqs.~\eqref{eq: holomorphic states},~\eqref{eq: holomorphic energy}, with $d=0$) being the ground state for $\alpha \rightarrow 0$ \cite{chou1991, nishida2007}.

The operator-state correspondence also teaches us about the structure of the energy spectrum for anyons in a harmonic trap, as it reveals that eigenstates can be organized into towers of states. For instance, $H_0-C+iD$ acts as a raising operator whose consecutive action on $\ket{\Psi_\mathcal{O}}$ increases the energy in steps of $2$, while the action of $P_i+iK_i$ (with $i\in\{x, y\}$ and $P_{i}$ and $K_{i}$ the total momentum and Galilean boost generators, respectively) increases the energy by 1, which is associated with the center-of-mass energy. These structures can all be seen as consistent with the exact energies shown in Eq.~\eqref{eq: 2-anyon exact} and Eq.~\eqref{eq: holomorphic energy}, but should also hold more generally for the harmonic trap problem.

\subsubsection{Anyonic coalescence behaviors}
\label{chap:coalesecence}
Lessons from NRCFT further teach us about the short-range behavior of hard-core anyons, a feature of relevance beyond the harmonic trap problem that offers important insights for our NNVMC implementation. Consider the real-space wavefunction of the state in Eq.~\eqref{eq:nrcft-state}, which can be written as a CFT correlator:
\begin{equation}
 \begin{aligned}
 &\Psi_{\mathcal O}(\br_1,\ldots,\br_N)\\
 \propto&\bra{0}\psi(0,\br_1)\cdots\psi(0,\br_N)e^{-H_0}\mathcal O^\dagger\ket{0}\\
 =&\bra{0}\psi(1,\br_1)\cdots\psi(1,\br_N)
             \mathcal O^\dagger\ket{0},
 \end{aligned}
 \label{eq:nrcft-wavefunction-correlator}
\end{equation}
where in the last line we used the imaginary-time evolution of the untrapped field $\psi(\tau,\br)=e^{\tau H_0}\psi(0,\br)e^{-\tau H_0}$. Consider $M\leq N$ particles approaching a common point $\bR$ at equal
imaginary time. Their positions are $\br_a=\bR+\delta\br_a$ for
$a=1,\ldots,M$. The operator product expansion (OPE) has the following form as $\delta\br_a \rightarrow 0 $:
\begin{equation}
 \begin{aligned}
 \psi(\tau,\br_1)\cdots
       \psi(\tau,\br_M) \sim
 \mathcal C_M(\{\delta\br_a\})\mathcal O_M(\tau,\bR)+\cdots,
 \end{aligned}
 \label{eq:nrcft-ope}
\end{equation}
where $\mathcal O_M$ denotes the $M$-particle primary with the \textit{lowest} scaling
dimension, and omitted terms contain
other primaries and descendants of higher scaling dimensions. Under a non-relativistic dilatation $(\tau,\br)\mapsto(b^2\tau,b\br)$, with $b>0$ and using Eq.~\eqref{eq: primary operator}, the fields $\psi$ and
$\mathcal O_M$ scale by a factor of $b^{-\Delta_\psi}$ and
$b^{-\Delta_{\mathcal O_M}}$, respectively \footnote{Concretely, under dilatation a primary operator transforms as $b^{\Delta_{\mathcal{O}}}\mathcal{O}(b^2\tau, b\br) = U_b\mathcal{O}(\tau,\br)U_b^{-1}$, with $U_b = e^{-i\ln(b)D}$, according to Eq.~\eqref{eq: primary operator}. Combining this with the covariance of the OPE in Eq.~\eqref{eq:nrcft-ope} implies the scaling relation in Eq.~\eqref{eq:nrcft-ope-scaling}.}.
Hence
\begin{equation}
 \mathcal C_M(\{b\,\delta\br_a\})
 =b^{\Delta_{\mathcal O_M}-M\Delta_\psi}
       \mathcal C_M(\{\delta\br_a\}),
 \label{eq:nrcft-ope-scaling}
\end{equation}
which implies $\mathcal C_M\propto\rho^{\Delta_{\mathcal O_M}-M\Delta_\psi}$,
with a shape-dependent prefactor, as the cluster at fixed shape
$\delta\br_a=\rho\,\mathbf u_a$ (with fixed $\bm{u}_{a}$ vectors with magnitude of order 1) shrinks with $\rho\to0^+$.

Notice that $\Delta_\psi=1$ and $\Delta_{\mathcal{O}_M}$ equals the $M$-anyon ground-state energy $E_M$ (by the assumption of $\Delta_{\mathcal{O}_M}$ being the smallest scaling dimension for an $M$-particle operator, see also Eq.~\eqref{eq:nrcft-state}). Combining the OPE with Eq.~\eqref{eq:nrcft-wavefunction-correlator}, we conclude that the generic $N$-anyon wavefunction would vanish, to the leading allowed order, as the following power law when $M$ anyons coalescence into a cluster with a characteristic size $\rho \rightarrow 0^+$:
\begin{equation}
 \Psi (\bR+\delta\br_{a=1,2,...,M};\br_{b=M+1,...,N})\ \sim\ \rho^{E_M-M}.
 \label{eq:nrcft-clustering-power}
\end{equation}
This is readily verified by the exact solution in Eq.~\eqref{eq: holomorphic states} and Eq.~\eqref{eq: holomorphic energy}, where the holomorphic state with energy $E^{(0)}_M$ is indeed the $M$-anyon ground state for small enough $\alpha$ given a fixed $M$.

Following the above discussion, it is straightforward to see that \textit{sub-leading} coalescence exponents for $M$ anyons are determined by the $M$-anyon \textit{excited-state} energies, as these are the scaling dimensions of the primary and descendant fields that are omitted in Eq.~\eqref{eq:nrcft-ope}. 
In particular, for the behavior when two anyons approach each other, combining Eqs.~\eqref{eq: two energies for 2 anyons} and \eqref{eq:nrcft-clustering-power}, we again obtain the two pair-coalescence exponents $\alpha$ and $2-\alpha$, which are the leading exponents that, respectively, increase and decrease linearly with the statistics parameter $\alpha$. 
As discussed in Sec.~\ref{chap:nn_general}, this is a crucial ingredient that we need to build into the neural network architecture for simulating anyon gases. In Sec.~\ref{sec: NNVMC results}, we further demonstrate that our trained anyonic quantum states can faithfully capture the predicted \textit{triple}-coalescence behavior.

\subsection{Anyonic atoms}
We introduce a second model Hamiltonian, which can be thought of as $N$ negatively-charged anyons along with a fixed positively-charged nucleus, which interact via Coulomb interactions in addition to the statistical gauge interaction. The purpose of introducing this model is twofold. First, this gives a simple model of interacting anyons which can be used as a benchmark for different variational ansatzes. Second, there has been recent interest in anyonic bound states appearing around impurities in fractional quantum anomalous Hall systems~\cite{li2026a, cui2026anyonicmoleculestrionlooking, Wagner_anyonimpurity_2026, adlong2026anyonsimpuritiesfractionalcharge}, and the model we study below may provide a conceptually simple starting point for understanding these impurity bound states and their generalizations. 

The Hamiltonian of $N$ anyons with charge $-1$ interacting with a nucleus of charge $Z$ at the origin is given by
\begin{equation}\label{eq: anyonic atom Hamiltonian}
    \begin{aligned}
      H_F &=
\frac{1}{2}\sum_{i=1}^N 
\left(-i\nabla_i+\mathbf a_i\right)^2 - \sum_{i=1}^{N}\frac{Z}{r_i} + \sum_{i < j}^{N} \frac{1}{\abs{\bm{r}_i - \bm{r}_j}}
    \end{aligned}
\end{equation}
in the fermion gauge. Here and throughout we adopt the Hartree units, defined by $\hbar=m=e=4\pi\epsilon_0=1$ (where $m$ and $-e$ are the mass and electric charge of the anyons).  While we deploy NNVMC to probe the ground state of this system for $N=2,3,4$ and generic statistics $\alpha$, see Sec.~\ref{sec: NNVMC results}, let us first remind the readers about the most trivial system of this kind, where $N=1$, and hence no statistical effects are present. 
This two-dimensional analogue of a hydrogen atom has the following spectrum \cite{kohn1955theory, zaslow1967two, PhysRevA.43.1186},
\begin{equation}
    E^{(n_r, m)}_1 = -\frac{Z^2}{2(n_r+\abs{m}+\frac{1}{2})^2},
\end{equation}
where $n_r \in \mathbb{Z}_0^+$ is the radial quantum number and $m\in\mathbb{Z}$ the angular momentum. 

In the non-interacting case, i.e., dropping the Coulomb repulsion between anyons, the ground state energy for a neutral bosonic atom with atomic number $Z=N$ is $-2N^3$, which serves as a strict lower bound for the anyonic atom energy for arbitrary $\alpha$, as both the statistical interaction (which leads to an effective exclusion effect away from the bosonic limit) and the Coulomb repulsion tend to raise the energy. As we will see in Sec.~\ref{sec: NNVMC results}, the anyonic atom energy generally increases as $\alpha$ increases between 0 (boson) and 1 (fermion). Another useful reference is the free-fermion atom (again take $N=Z)$, where the ground state energy is $-80/9$ and $-22$, for $N=$ 2 and 3, respectively. 

\section{Neural-Network Ansatzes and Their Challenges}
\label{chap:nn_general}
In this section, we provide a brief overview of neural-network ansatze for variational Monte Carlo (VMC) calculations.
In part \ref{chap:ferminet}, we explain the FermiNet architecture, which we use throughout this study, along with the main concepts common in essentially all fermionic neural-network ansatze.
Next, in part \ref{chap:nn_for_anyons}, we discuss extra restrictions that the neural-network ansatze must
satisfy if they are to be used to study fundamental anyonic systems, related to the non-analytic short-range behavior of anyons.
Finally, in part \ref{chap:jastrow_ansatze}, we argue that the most natural way of enforcing these short-range conditions, using a Jastrow factor over pairs multiplied by a smooth network, is insufficient.
Then, in section \ref{chap:architecture}, we present the ansatz used in this study, and the modifications made to ensure the correct short-range structure.

\subsection{FermiNet}
\label{chap:ferminet}
In this section, we provide an overview of NNVMC, and explain the FermiNet architecture, which we use throughout this study.
The FermiNet \cite{pfau2020ab, spencer2020better}, and related architectures \cite{hermann2020, glehn2023, schatzle2023deepqmc} is a fermionic neural network ansatz.
In other words, the FermiNet is a function which takes as inputs the coordinates and (generally, although in this work we only have a single spin species) the spins of fermions $\{\bm{r}_{i}, \sigma_{i}\}$, and outputs a prediction for the value of the wavefunction at that configuration.
The form of the ansatz is a neural-network: the configuration is used to create ``features" which are then passed through dense linear layers followed by non-linear activation functions.
The weights and biases that define the neural network ansatz are then optimized to minimize the expectation value of the Hamiltonian, following the VMC algorithm \cite{foulkes2001quantum}.
In order to describe a system of fermions, the network must be totally antisymmetric with respect to exchange of fermion coordinates (from now on, we assume a single spin species).
This is done by constructing generalized Slater determinants 
\begin{align}
    \psi(\bm{r}_{1}, \dots, \bm{r}_{N}) &= \sum_{k=1}^{N_{\text{dets}}}\det_{ij}[\phi_{i}^{k}(\bm{r}_{j}, \{\bm{r}_{/j}\})],
\end{align}
where $\{\bm{r}_{/j}\}$ means the set of fermion positions other than that of fermion $j$.
We note that these generalized determinants are significantly more expressive than traditional Slater determinants \cite{carleo2026fixed}.
If the positions of two fermions $i^{\prime}$ and $j^{\prime}$ are exchanged, then columns $i^{\prime}$ and $j^{\prime}$ of $\phi_{i}^{k}(\bm{r}_{j}, \{\bm{r}_{/j}\})$ (for a fixed $k$) are exchanged.
As determinants are odd under exchanges of columns, every determinant in the sum transforms by a negative sign, making the total wavefunction odd under exchanges of fermion positions.
What has been discussed so far is common for virtually all NNVMC approaches and architectures (with the exception of other antisymmetrization methods such as Pfaffians).
The difference lies in the way the ``orbitals" $\phi^{k}_{i}(\bm{r}_{j}, \{\bm{r}_{/j}\})$ are represented. 
The FermiNet represents the generalized orbitals as follows.

We start by defining a set of features, or input, vectors.
The FermiNet contains two ``streams", a single-particle stream and a two-particle stream.
The input features for the one $\bm{h}_{i}^{0}$ and two $\bm{h}_{ij}^{0}$ particle streams are
\begin{align}
    \bm{h}_{i}^{0} &= (\bm{r}_{i} - \bm{R}_{I}, |\bm{r}_{i} - \bm{R}_{I}|\quad \forall I),\\
    \bm{h}_{ij}^{0} &= (\bm{r}_{i} - \bm{r}_{j}, |\bm{r}_{i} - \bm{r}_{j}|),
\end{align}
where $i,j$ label fermions, $I$ labels the atoms in the system and $\bm{R}_{I}$ is the position of atom $I$.
In the harmonic oscillator case, the location of the center of the well replaces the atomic coordinates $\bm{R}_{I}$.
The superscript refers to the layer of the network, with layer ``0" meaning the inputs.
From the $l$th layer, one can compute the $(l+1)$th layer in the following way,
\begin{align}
    \bm{g}^{l} &= \frac{1}{N}\sum_{i}\bm{h}_{i}^{l},\\
    \bm{g}_{i}^{l} &= \frac{1}{N}\sum_{j}\bm{h}_{ij}^{l},\\
    \bm{f}_{i}^{l} &= (\bm{h}_{i}^{l}, \bm{g}^{l}, \bm{g}_{i}^{l}),\\
    \bm{h}_{i}^{l+1} &= \tanh(\bm{V}_{\text{FN}}^{l}\bm{f}_{i}^{l} + \bm{b}_{\text{FN}}^{l}) + \bm{h}_{i}^{l},\\
    \bm{h}_{ij}^{l+1} &= \tanh(\bm{W}_{\text{FN}}^{l}\bm{h}_{ij}^{l} + \bm{c}_{\text{FN}}^{l}) + \bm{h}_{ij}^{l},
\end{align}
where $\bm{V}_{\text{FN}}^{l}$, $\bm{W}_{\text{FN}}^{l}$, $\bm{b}_{\text{FN}}^{l}$ and $\bm{c}_{\text{FN}}^{l}$ are learnable parameters.
We use the FN subscript to denote that these are FermiNet parameters; our network will have additional parameters.
Finally, the orbitals are defined by
\begin{align}
    \phi_{i}^{k}(\bm{r}_{j}, \{\bm{r}_{/j}\}) &= (\bm{w}_{\text{FN}, i}^{k}\cdot \bm{h}_{j}^{L})\xi_{i}(\bm{r}_{j}),
\end{align}
where $\bm{w}_{\text{FN}, i}^{k}$ are learnable parameters and $\xi_{i}$ are envelope functions used to meet the boundary conditions $\psi(|\bm{r}|\to\infty)\to 0$.

\subsection{Neural networks for anyons}
\label{chap:nn_for_anyons}
In the fermion gauge, the behavior of anyonic wavefunctions with statistics parameter $\alpha$ as two particles $i$ and $j$ approach each other can be any of 
\begin{align}
    \chi_{F}^{(n)}(\bm{r}_{ij}) &= r_{ij}^{|\alpha + 2n|}e^{i(2n+1)\theta_{ij}}\;\;\;(n \in\mathbb{Z}).
    \label{eq:channels}
\end{align}
This is both explained by our discussion in Sec. \ref{sec: theory and previous works} and also established by an analysis of the $N=2$ anyon problem as detailed in Appendix \ref{chap:short_range}. Notice that in the presence of anyon-anyon Coulomb interaction, the above short-distance behavior should be further multiplied by a factor $(1+\frac{r}{2|\alpha+2n|+1})$ as interactions modify the subleading short-range behavior. 

The primary goal in designing our variational states is to exactly satisfy the non-analytic short-distance behavior while retaining the robust expressivity of architectures such as FermiNet. 
Failure to represent exactly this structure results in a divergent Monte Carlo variance of the local energy, see Appendix \ref{chap:variance}. 
It is possible for the network to learn to approximate some non-analytic behavior; indeed, NNVMC studies of 2D electrons with Coulomb interactions successfully take this approach~\cite{nazaryan2026a, avdoshkin2026, geier2025, gaggioli2026}. 
We find this approach to be inadequate for our systems, likely due to the $r^{-2}$ divergence of the interaction rather than a Coulombic $r^{-1}$. 
This results in very poorly optimized wavefunctions. 

The simplest way of enforcing the necessary short-range condition is to use a Jastrow factor over pairs, $\psi_{\text{Jastrow}} = \prod_{i<j} \mathcal{J}(\bm{r}_{i}, \bm{r}_{j}; \alpha) \times \phi_{\text{NN}}(\bm{r}_{1}, \dots, \bm{r}_{N})$, where $\phi_{\text{NN}}$ is obtained from the output of a smooth neural-network, and $\mathcal{J}$ is constructed from the allowed channels in Eq.~\eqref{eq:channels}.
However, as discussed in Section \ref{chap:jastrow_ansatze}, any ansatz written in this way will either encounter the infinite variance problem, or not be able to represent a general state near $\alpha = 1$.

Instead, we propose a general method of modifying the orbitals of a generalized Slater determinant, whose bulk neural network is given by the FermiNet architecture, to enforce a desired short-distance behavior.
The full ansatz takes the form of a sum of a determinant-based function ($\psi_{\rm Det}$), and a Jastrow-Bose function ($\psi_{\rm DS}$) gauge-transformed into fermion gauge.
The bulk of $\psi_{\rm DS}$ is a Deep Set \cite{zaheer2017deep} with an additive pooling layer. 
We find that $\psi_{\rm Det}$ is generally effective for capturing nonlinear branches for the harmonic trap problem, while $\psi_{\rm DS}$ is effective for capturing the linear holomorphic branch near the bosonic end, and altogether our full ansatz can learn the ground state of anyon gas for a generic statistics parameter $\alpha$.
The form of the ansatz as presented is to be used in the fermion gauge, although we stress that every aspect of our calculation, including stochastic estimation of observables, is gauge-invariant and can be equally regarded as an evaluation of a multi-valued anyonic wavefunction.
We describe the determinant-based ansatz in section \ref{chap:determinant}, and the Jastrow-Bose ansatz in section \ref{chap:Deep Set}.
Details about the optimization and choice of hyperparameters can be found in Appendix \ref{chap:optim}.
We call our new architecture the AnyNet.

\subsection{Failure of Jastrow-Based Ansatze}
\label{chap:jastrow_ansatze}
A very natural way of enforcing the correct short-range behavior is to include it as a Jastrow factor over pairs of particles,
\begin{align}
    \psi_{\text{Jastrow}} = \prod_{i<j} \mathcal{J}(\bm{r}_{i}, \bm{r}_{j}; \alpha) \times \phi_{\text{NN}}(\bm{r}_{1}, \dots, \bm{r}_{N}),\label{eq:jastrow_ansatz}
\end{align}
where $\phi_{\text{NN}}$ is a smooth neural network, taken to be either a general symmetric function (with an antisymmetric $\mathcal{J}$) or a general antisymmetric function (with a symmetric $\mathcal{J}$). 
However, we argue below that no such ansatz can satisfactorily be used to study non-interacting anyons, whose short-range behavior is given by Eq.~\eqref{eq:channels}.

\begin{figure}[h!]
    \centering
    \includegraphics[width=\columnwidth]{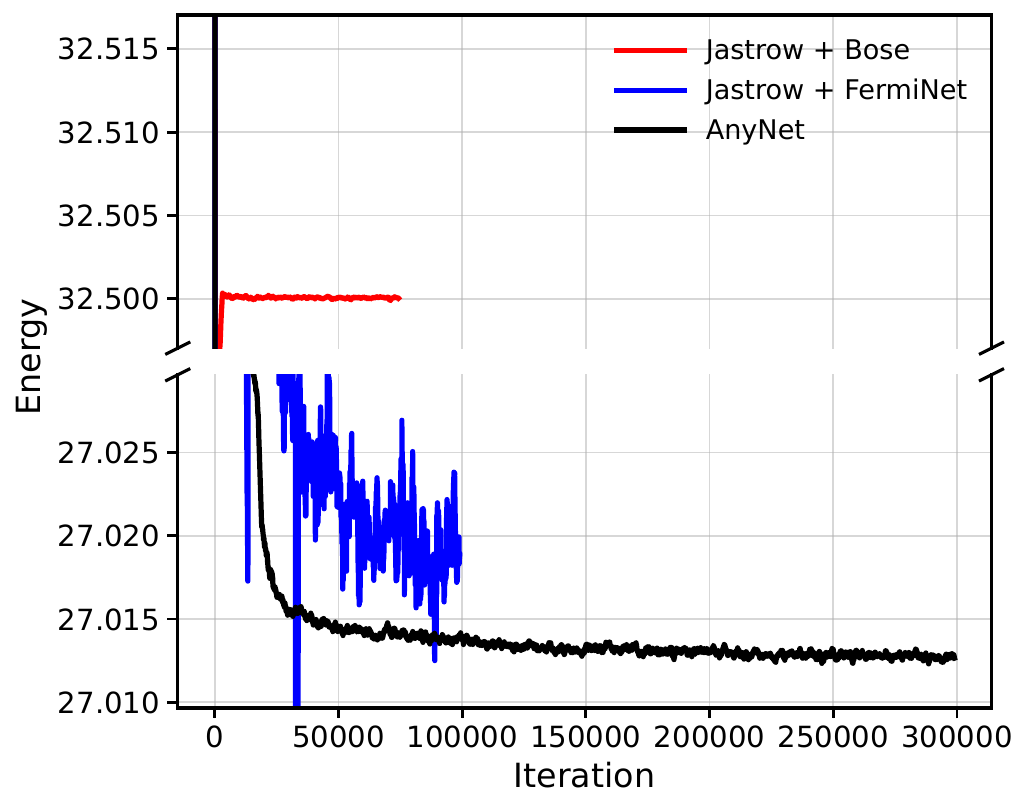}
    \caption{Learning curve for $N = 10$ anyons at exchange parameter $\alpha = 0.5$ (semions).
    The Jastrow + FermiNet ansatz has a divergent variance problem, which makes it practically challenging to optimize.
    In contrast, the Jastrow + Bose ansatz is unable to represent the ground state and gets stuck in the holomorphic branch.}
    \label{fig:learning_curve_comparison}
\end{figure}

Consider first the case where $\phi_{\text{NN}}$ is an antisymmetric function.
For generic $\alpha$, the allowed channels in Eq.~\eqref{eq:channels} have a one-to-one correspondence between their radial and angular behavior. 
By this, we mean that given a radial short-range behavior, there is a unique angular function that is a solution of the Schr\"odinger equation.
For example, suppose the radial part of the wavefunction as two anyons approach each other is $r^{\alpha}$.
Then, there is a unique angular distribution, $\Theta(\theta)$, such that $r^{\alpha}\Theta(\theta)$ is a solution to the Schr\"odinger equation near $r = 0$.
In this case, the required angular distribution is $\Theta(\theta)=\exp(i\theta)$.
As demonstrated in Appendix \ref{chap:variance}, any state that does not exactly solve the Schr\"odinger equation near coincidences has a \textit{local energy divergence} which causes the variance of the Monte Carlo estimate of the energy to diverge.
The problem with a generic antisymmetric function is that as two particles approach each other, there are two equally valid angular configurations.
For example, if the radial distribution is $r^{\alpha}$, (with $\mathcal{J} = r^{\alpha -1}$), then $\phi_{\text{NN}}$ has a Taylor expansion of $r(a^{+}e^{i\theta} + a^{-}e^{-i\theta})$, which must satisfy $a^{-}\equiv 0$ to keep the variance of the energy finite. 
However, $a^{-}$ depends on the position of all anyons except the two that are near coincidence, and $a^{-}$ must be fine-tuned to be exactly 0 for all possible configurations.
While a network could, in principle, represent such a state, the initial, random network does not satisfy this condition.
Therefore, from the start of training, the estimated energies and energy gradients have a divergent variance, which in practice prohibits training towards the required $a^{-}\equiv 0$.
The antisymmetric $\phi_{\text{NN}}$ case therefore has a trainability issue, not a representability issue.
An example learning curve is shown in Figure \ref{fig:learning_curve_comparison}.

Consider now the case of $\phi_{\text{NN}}$ being a fully symmetric and smooth network.
As a generic symmetric function does not vanish whenever two particles coincide, the full radial power must be encoded into the Jastrow factor.
Hence, the most general Jastrow factor (excluding terms that can be absorbed into the smooth network $\phi_{\text{NN}}$) is
\begin{align}
    \mathcal{J}(\bm{r}_{i}, \bm{r}_{j}; \alpha) &= \sum_{k}c_{k}\chi_{F}^{(k)}(\bm{r}_{i} - \bm{r}_{j}).
\end{align}
The problem is that if two particles are within a distance $\epsilon$ of each other, then $\chi_{F}^{(k)}(\bm{r}_{i} - \bm{r}_{j}) \sim \mathcal{O}(\epsilon^{\alpha})$ and hence $\mathcal{J}(\bm{r}_{i}, \bm{r}_{j}; \alpha)\sim \mathcal{O}(\epsilon^{\alpha})$.
Consider now what happens when three particles approach each other. 
There are 3 terms in the Jastrow product that scale like $\sim \mathcal{O}(\epsilon^{\alpha})$, so the full wavefunction scales as $\psi_{\text{Jastrow}}\sim \mathcal{O}(\epsilon^{3\alpha})$.
However, the generic behavior for three fermions ($\alpha = 1$) approaching each other in 2D is $\sim \mathcal{O}(\epsilon^{2})$, which is clear both from Eq.~\eqref{eq:nrcft-clustering-power} (and recall $E_3=5$) or by a direct inspection of the free-fermion ground-state wavefunction in a harmonic trap. 
Therefore, an ansatz of the form of Eq.~\eqref{eq:jastrow_ansatz} (with $\phi_{\text{NN}}$ fully symmetric) cannot represent the three-body structure of fermions, as it decays too quickly.
More generally, the Jastrow-based ansatz only contains total relative angular momentum of 3 and above, while a generic fermionic state contains components of total relative angular momentum 2.
Therefore, the Jastrow-based ansatz is also unable to capture the correct phase structure around triplets of particles.
This obstruction is only a problem for $N \ge 3$, meaning that for $N = 2$, a Jastrow-based ansatz is able to represent the ground state wavefunction, which we have verified computationally. Empirically, we also find that at intermediate $\alpha$, this ansatz often converges to the linear branch in systems where the linear branch is not the ground state.
An example is $N = 10$, $\alpha = 0.5$, where the Jastrow + Bose ansatz (Eq.~\eqref{eq:jastrow_ansatz} with a symmetric $\phi_{\text{NN}}$ and an antisymmetric $\mathcal{J}$) converges to a state with energy 32.5, consistent with the holomorphic state, while we find the ground state at an energy of 27.01276(5) with the AnyNet, discussed in Section \ref{chap:architecture}.
This comparison is shown in Figure~\ref{fig:learning_curve_comparison}. 

\section{The AnyNet Architecture}
\label{chap:architecture}
\begin{figure*}
    \centering
    \includegraphics[width=0.99\linewidth]{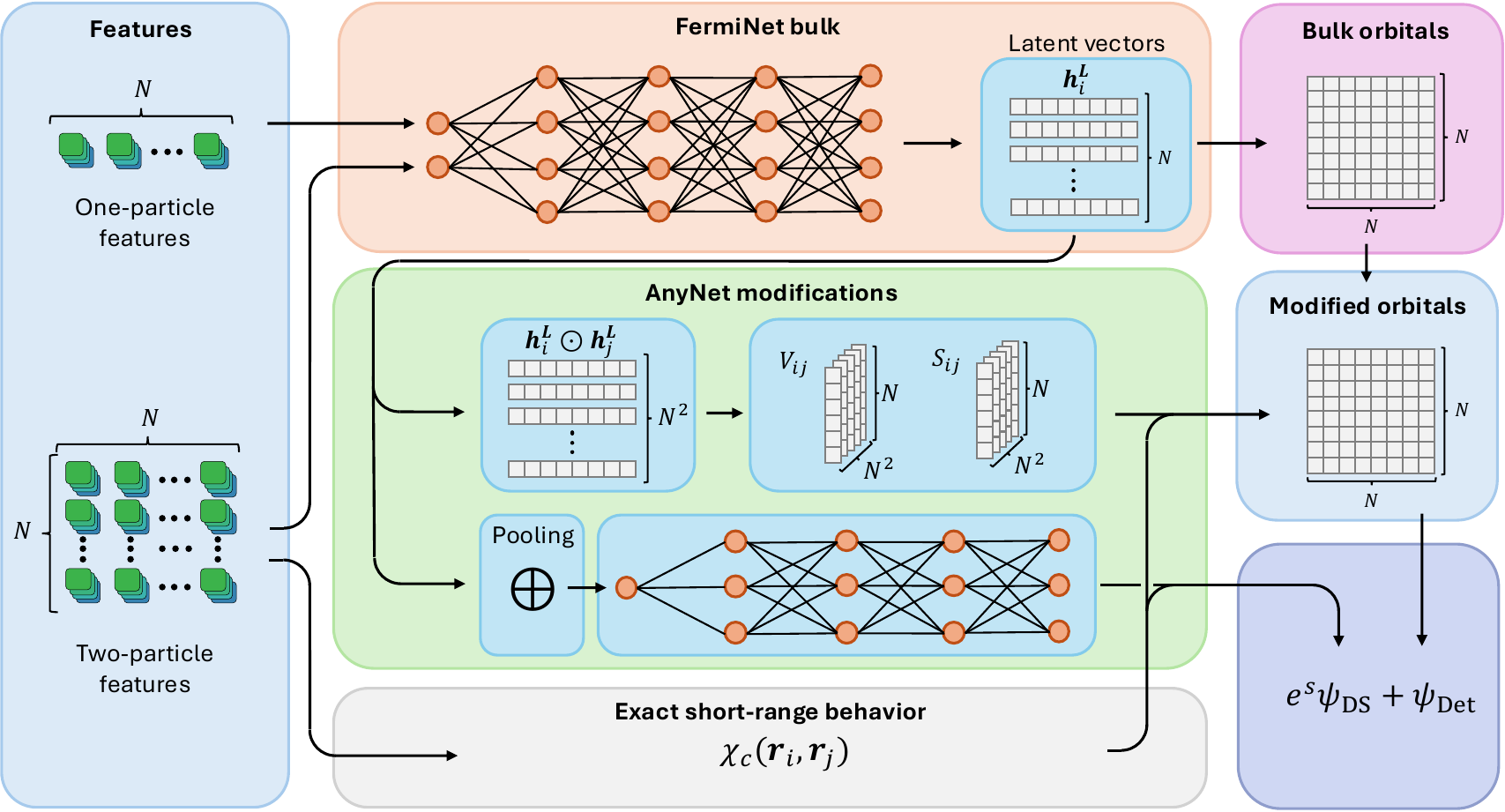}
    \caption{
    Schematic of the AnyNet network.
    We start by obtaining single-particle features, such as the position of the anyons, $\{\bm{r}_{i}\}$, and two-particle features, such as the relative displacements between pairs $\{\bm{r}_{i} - \bm{r}_{j}\}$.
    These features are then fed into the FermiNet architecture, from which the latent vectors $\bm{h}_{i}^{L}$ are obtained.
    The latent vectors are used to construct ordinary (bulk) orbitals, and the $V_{ij}$ and $S_{ij}$ tensors, which describe the shape of the wavefunction near anyon coincidences.
    The latent vectors are also pooled (averaged) and fed to a second small MLP, a method usually called Deep Set in machine learning.
    The bulk orbitals are then modified with the $V_{ij}$ and $S_{ij}$ tensors to encode the short-range behavior analytically.
    The wavefunction is evaluated by taking a weighted sum between $\psi_{\rm DS}$ (the Deep Set output times a Jastrow), and the determinant of the modified Slater matrix ($\psi_{\rm Det}$).}
    \label{fig:architecture}
\end{figure*}

In this section, we present the AnyNet architecture, a method for modifying a determinant-based neural network ansatz to analytically represent any desired short-range behavior.
We note that although we use FermiNet as an example throughout, any neural network ansatz which uses the latent vectors to construct generalized single-particle orbitals, such as the Psiformer \cite{glehn2023}, can be modified in the same way.
The pairwise short-range behavior that is enforced can also be chosen flexibly; thus the method is not specific to the anyon problem and we expect our construction to be useful for a wide range of exotic quantum systems.
A schematic representation of our ansatz can be found in Figure \ref{fig:architecture}. 

\subsection{The Determinant architecture}
\label{chap:determinant}
The determinant component of the ansatz is a modification on top of the FermiNet architecture which we have reviewed in Sec.~\ref{chap:ferminet}.
Consider a determinant of $N$ orbitals $\phi_i(\bm{r}_{j}, \{\bm{r}_{/j}\})$. 
As before, the $\{\bm{r}_{/j}\}$ notation means that the orbitals depend on $\bm{r}_{j}$, as well as all other fermions in a permutation-equivariant manner. Define
\begin{equation}
    \psi_{\rm Det}(\bm{r}_{1}, \dots, \bm{r}_N) = \sum_{k}^{N_{\text{dets}}}\text{det}_{ij}[\phi_{i}^{k}(\bm{r}_{j}, \{\bm{r}_{/j}\})].
\end{equation}
Let $\phi_{i}^{k}(\bm{r}_{j}, \{\bm{r}_{/j}\})\equiv[\bm{\Phi}^{k}(\bm{r}_j, \{\bm{r}_{/j}\})]_{i}$, in other words, $\bm{\Phi}^{k}(\bm{r}_{j}, \{\bm{r}_{/j}\})$ is the $j$th column of the $k$th determinant. 
As particles $i$ and $j$ approach each other, we demand that the $i$'th and $j$'th column smoothly deform into $\bm{f}^{k}(\bm{r}_i, \bm{r}_j, \{\bm{r}_{/(i,j)}\})$ and $\bm{f}^{k}(\bm{r}_j, \bm{r}_i, \{\bm{r}_{/(j,i)}\})$, respectively, with the following properties:
\begin{align}
    \bm{f}^{k}(\bm{r}_i, \bm{r}_j) &= \bm{S}^{k}(\bm{r}_i, \bm{r}_j) + \bm{\Delta}^{k}(\bm{r}_i, \bm{r}_j),\\
    \bm{\Delta}^{k}(\bm{r}_i, \bm{r}_j) &= \sum_{c}\chi_{c}(\bm{r}_i - \bm{r}_j) \bm{V}^{k}_{c}(\bm{r}_i, \bm{r}_j),\\
    \bm{S}^{k}(\bm{r}_i, \bm{r}_j) &= \bm{S}^{k}(\bm{r}_j, \bm{r}_i),\\
    \bm{V}^{k}_{c}(\bm{r}_i, \bm{r}_j) &= \bm{V}^{k}_{c}(\bm{r}_j, \bm{r}_i),\\
    \chi_{c}(\bm{r}) &= - \chi_{c}(-\bm{r}),
\end{align}
where $\bm{S}^{k}$ and $\bm{V}_{c}^{k}$ are learnable functions whose parametrization is discussed later, and we have dropped the $\{\bm{r}_{/(i,j)}\}$ arguments to avoid clutter.
This decomposition is chosen specifically so that the antisymmetric scalar functions $\chi_{c}(\bm{r}_i - \bm{r}_j)$ control the short-distance behavior as $\bm{r}_i - \bm{r}_j \rightarrow 0$.
Here, $c \in \mathbb{Z}$ labels the allowed short range channels described in Appendix~\ref{chap:short_range}, and Eq.~\eqref{eq:channels}.
Note that while it would be natural to write $\chi_{c}(\bm{r}_{i}, \bm{r}_{j})=r_{ij}^{|\alpha + 2c|}\exp[i(1+2c)\theta_{ij}]$, we found that this can grow very large, and the $\chi_{c}$ functions end up dominating the form of the wavefunction. 
This is particularly problematic for large $N$.
Instead, we adopt 
\begin{align}\label{eq: chi_c}
    \chi_{c}(\bm{r}_{i}, \bm{r}_{j})&=\tanh(r_{ij})^{|\alpha + 2c|}e^{i(1+2c)\theta_{ij}},
\end{align}
which saturates to 1 at large separation. When anyons are also interacting via the Coulomb potential (e.g. for our study of the anyon atom in Sec.~\ref{chap:anyonic_atoms}), the above short-distance channel function is further multiplied by a factor of $\left(1+\frac{\tanh{r_{ij}}}{2|\alpha+2c|+1}\right)$, as the Coulomb interaction modifies the subleading short-range behavior (Appendix \ref{chap:short_range}).

Consider, as an example, a single channel ($c = 0$) and two particles, $i=1$ and $j=2$, coalescing. 
Dropping the $k$ index for concreteness, and using the notation $(\cdot)_{ij}\equiv (\cdot)(\bm{r}_{i}, \bm{r}_{j})$, 
\begin{equation}
    \begin{aligned}
        &\psi_{\rm Det}(\bm{r}_1, \bm{r}_2, \ldots, \bm{r}_N)\\
        &= \det \begin{pmatrix}
            \bm{S}_{12} + \bm{\Delta}_{12},
            &
            \bm{S}_{12} - \bm{\Delta}_{12},
            &
            \bm{\Phi}_3,
            &
            \ldots,
            &
            \bm{\Phi}_N
        \end{pmatrix}
        \\
         &\propto \det \begin{pmatrix}
             \bm{\Delta}_{12},
            &
            \bm{S}_{12}, 
            &
            \bm{\Phi}_3,
            &
            \ldots,
            &
            \bm{\Phi}_N
            \end{pmatrix}\\
            &= \chi(\bm{r}_1 - \bm{r}_2)\det \begin{pmatrix}
             \bm{V}_{12},
            &
            \bm{S}_{12} ,
            &
            \bm{\Phi}_3,
            &
            \ldots,
            &
            \bm{\Phi}_N
            \end{pmatrix}.
    \end{aligned}
\end{equation}
This structure is enforced by constructing the columns as
\begin{align}
    &\bm{\Phi}^{k}(\bm{r}_{i}) = v_{i}\bm{\Lambda}^{k}(\bm{r}_{i}) +\sum_{j\ne i}w_{ij} \bm{f}^{k}(\bm{r}_{i}, \bm{r}_{j}),\\
    &v_{i} = \frac{1}{1+\sum_{j \ne i}u_{ij}},\;\;w_{ij} = u_{ij} v_i,\;\;u_{ij} = \left(\frac{r_c}{r_{ij}}\right)^{2},
\end{align}
where $r_{c}$ is a constant, and $\bm{\Lambda}^{k}(\bm{r}_{i})$ are the FermiNet generalized single-particle orbitals, but they do not play a role when two particles approach each other.
The weights obey the condition that as two particles $i$ and $j$ approach each other, $v_{i}, v_{j} \to 0$, $w_{/(i,j)}\to 0$, $w_{ij} \to 1$, guaranteeing the desired form of the determinant.

The symmetric functions $\bm{S}^{k}$, $\bm{V}^{k}_{c}$ are obtained as bilinear products of the last layer of the FermiNet, following Ref.~\cite{lou2024neural}.
Concretely, let the latent vector for particle $i$ be $\bm{h}_{i}^{L}$, of dimension $d$. 
The bulk of the network, from which the $\bm{h}_{i}^{L}$ latent vectors are obtained, is that of the FermiNet architecture \cite{ferminet_github, pfau2020ab, spencer2020better}.
To represent the symmetric (under exchange of $i$ and $j$) function $\bm{S}^{k}(\bm{r}_{i}, \bm{r}_{j})$, with $\dim[\bm{S}^{k}(\bm{r}_{i}, \bm{r}_{j})] = N$, we first find the Hadamard product (element-wise multiplication $\odot$) of the latent vectors for every pair, $\bm{K}_{ij} = \bm{h}_{i} \odot \bm{h}_{j}$, and apply a linear transformation
\begin{align}
    \bm{S}^{k}(\bm{r}_{i}, \bm{r}_{j}) &= [W_{\bm{S}}]^{k} \bm{K}_{ij},
\end{align}
where $[W_{\bm{S}}]^{k}$ are learnable matrices of dimension $N$ by $d$.
The process for representing the $\bm{V}_{c}^{k}$ is equivalent, but with a different learnable linear transformation defined by parameters $[W_{\bm{V}}]_{c}^{k}$.

\subsection{The Deep Set architecture}
\label{chap:Deep Set}
Empirically, we find that the determinant architecture alone is not sufficient to describe the holomorphic branch (for almost-bosonic particles).
The remedy is to add to our ansatz a term that accurately represents the holomorphic branch.
This is done by designing an accurate, fully symmetric (bosonic) neural network wavefunction, and applying the necessary gauge transformation to transform into the fermion gauge.

The fully symmetric bulk is a simple Deep Set \cite{zaheer2017deep} with an additive pooling layer.
The pooling layer is a simple additive one; we define an average latent vector $\bar{\bm{h}} = \frac{1}{N}\sum_{i}\bm{h}_{i}^{L}$, where, as before, $\bm{h}_{i}^{L}$ is the output of the last layer of the FermiNet applied to particle $i$.
$\bar{\bm{h}}$ is explicitly permutation invariant, so any function of $\bar{\bm{h}}$ must also be permutation invariant.
$\bar{\bm{h}}$ is used as the input feature to a second multi-layer perceptron, composed of two linear layers with tanh activation functions and skip connections (an identity path for information to be transferred to the following layer).
Our implementation follows
\begin{align}
    \bm{\mathfrak{g}}^{0} &= \bar{\bm{h}},\\
    \bm{\mathfrak{g}}^{l + 1} &= \tanh(W_{D}^{l}\bm{\mathfrak{g}}^{l} + \bm{b}_{D}^{l}) + \bm{\mathfrak{g}}^{l} \quad (l+1 < L_{D}),\\
    \bm{\mathfrak{g}}^{L_{D}} &= W_{D}^{L_{D} - 1}\bm{\mathfrak{g}}^{L_{D} - 1}
\end{align}
where $\bm{\mathfrak{g}}^{l}$ is the activation in layer $l$ of the Deep Set, $W_{D}^{l}$, $\bm{b}_{D}^{l}$ are learnable parameters, and $L_{D} = 2$ is the number of layers in the Deep Set.
To ensure the ansatz has the correct dynamical range, we have found it beneficial for the network to return $2N$ real outputs, $\dim(\bm{\mathfrak{g}}^{L_{D}}) = 2N$. 
These outputs are then combined into $N$ complex outputs $\widetilde{\mathfrak{g}}_{k} = \mathfrak{g}^{L_{D}}_{2k} + i\mathfrak{g}^{L_{D}}_{2k+1}$.
The Jastrow-Deep Set part of the ansatz is then written as
\begin{align}
    \psi_{\rm DS}(\bm{r}_{1}, \dots, \bm{r}_{N}) &= \left[\prod_{i<j}\chi_{c = 0}(\bm{r}_{i} - \bm{r}_{j})\right]\left[\prod_{k}\widetilde{\mathfrak{g}}_{k}\right],\label{eq:bose_jastrow}
\end{align}
where only the $c=0$ term (see Eq.~\eqref{eq: chi_c}) is needed for the Jastrow factor, as this is the only channel present in the holomorphic branch. 

\begin{table*}[!t]
\centering
\begin{tabular}{|c|c c c c c c c c|}
  \hline
  $\alpha$ & 0.0 & 0.2 & 0.4 & 0.6 & 0.7 & 0.8 & 0.9 & 1.0\\
  \hline
  ED & 4.0000000(0) & 5.200000(0) & 6.400000(0) & 7.41048(2)\phantom{0} & 7.47728(4)\phantom{0} & 7.59887(1)\phantom{0} & 7.774072(7) & 8.000000(0)\\
  AnyNet & 4.0000005(4) & 5.200002(2) &
  6.400001(4) & 7.410489(7) & 7.477283(5) & 7.598884(4) & 7.774097(4) & 8.000017(4)\\
  FermiNet & 5.033(4)\phantom{0000} & 5.464(2)\phantom{000} & 6.448(2)\phantom{000} & 7.4146(5)\phantom{00} & 7.4795(2)\phantom{00} & 7.5995(2)\phantom{00} & 7.77428(5)\phantom{0} & 8.000001(1)\\
  \hline
  $\Delta\ [10^{-6}]$ & 0.5(4) & 2.0(20) & 1.0(40) & 9.1(23) & 2.8(37) & 12.9(13) & 25.5(8) & 17.0(4)\\
  \hline
\end{tabular}
\caption{$N = 4$ energy benchmarks with the bare FermiNet, and exact diagonalization (ED) extrapolated to the complete basis limit for a range of $\alpha$.
$\Delta$ corresponds to the energy difference, in $10^{-6}$ energy units, between the ED results and the AnyNet.
At $\alpha \lesssim 0.55$, the ground state is always in the linear branch of exactly known holomorphic states, which can be easily represented and we instead concentrate on the non-trivial, non-linear branch.
ED calculations for anyons are very challenging, as the vector potential, when expanded, induces a 3-body interaction, making the computation costly.
Therefore, the available ED data is only for very small bases, and the extrapolation to the infinite basis limit results in the quoted error bars.
For more information on the extrapolation to the infinite basis limit, see Appendix \ref{chap:ED_extrapolation} and Figure \ref{fig:ED_extrapolation}.
\label{table:energies}
}
\end{table*}

As $\chi_{c}$ are antisymmetric functions, $\psi_{\rm DS}$ is fully antisymmetric, and is therefore a valid wavefunction for the anyons in the fermion gauge.
The overall ansatz is written as 
\begin{align}
    \Psi_F = ( \psi_{\rm Det}+e^{s}\psi_{\rm DS})\prod_{i}\xi(\bm{r}_{i}),
\end{align}
where $\xi(\bm{r})$ is an envelope used to analytically satisfy the boundary conditions at large $|\bm{r}|$.
For the harmonic oscillator problem, $\xi(\bm{r}) = \exp(-r^{2}/2)$ was used, and for the anyonic atoms $\xi(\bm{r}) = \exp(-\sigma r)$, with $\sigma$ being a learnable parameter.
We remark that $s$, which parametrizes the weight between the two architectures is one of the training parameters, and is initialized to $-N$ as we find that the network can efficiently ``turn on" the Deep Set part of the ansatz but, at times, struggles to reduce its weight when it is not needed.

\section{Results}\label{sec: NNVMC results}

\subsection{Energy spectrum and angular momentum}
\label{chap:ED_comparison}
Exact diagonalization (ED) of the free anyon gas in a harmonic potential has been performed for $N=3$~\cite{sporre1991, murthy1991} and $N=4$~\cite{sporre1992}. This should be constrasted to more conventional interacting fermionic systems in a 2D harmonic trap, which have been studied up to $N=8$~\cite{rontani2006} with full configuration interaction (FCI), and up to $ N=20$ with coupled cluster with single, double, and perturbative triple excitations (CCSD(T)) \cite{pedersen2011ab}.
The anyonic system presents more challenges due to both the highly non-analytic nature of the wavefunction, which makes errors from Hilbert space truncation more severe, as well as the three-body interaction generated by the statistical gauge field. The $N=3$ and $N=4$ numerical studies are aided by specialized coordinate transformations that do not generalize to larger systems, as well as a similarity transformation that removes the three-body interaction at the cost of making the Hamiltonian non-Hermitian.
We compare our NNVMC results to the $N = 4$ ED energies extrapolated to the infinite basis limit.
Both methods agree to a few parts in a million.
We also demonstrate the failure of the default FermiNet at describing anyon systems, even at $N = 4$, particularly for small $\alpha$.
The results are presented in Table \ref{table:energies}. Further details of the ED extrapolation are in Appendix~\ref{chap:ED_extrapolation}; we mention here that the extrapolation is exceptionally challenging due to the high sensitivity of the ground state energy to the Hilbert space truncation, as well as the lack of a variational principle due to the non-Hermiticity of the Hamiltonian. These difficulties further motivate our numerical methods.

\begin{figure*}[t]
    \centering
    \includegraphics[width=\textwidth]{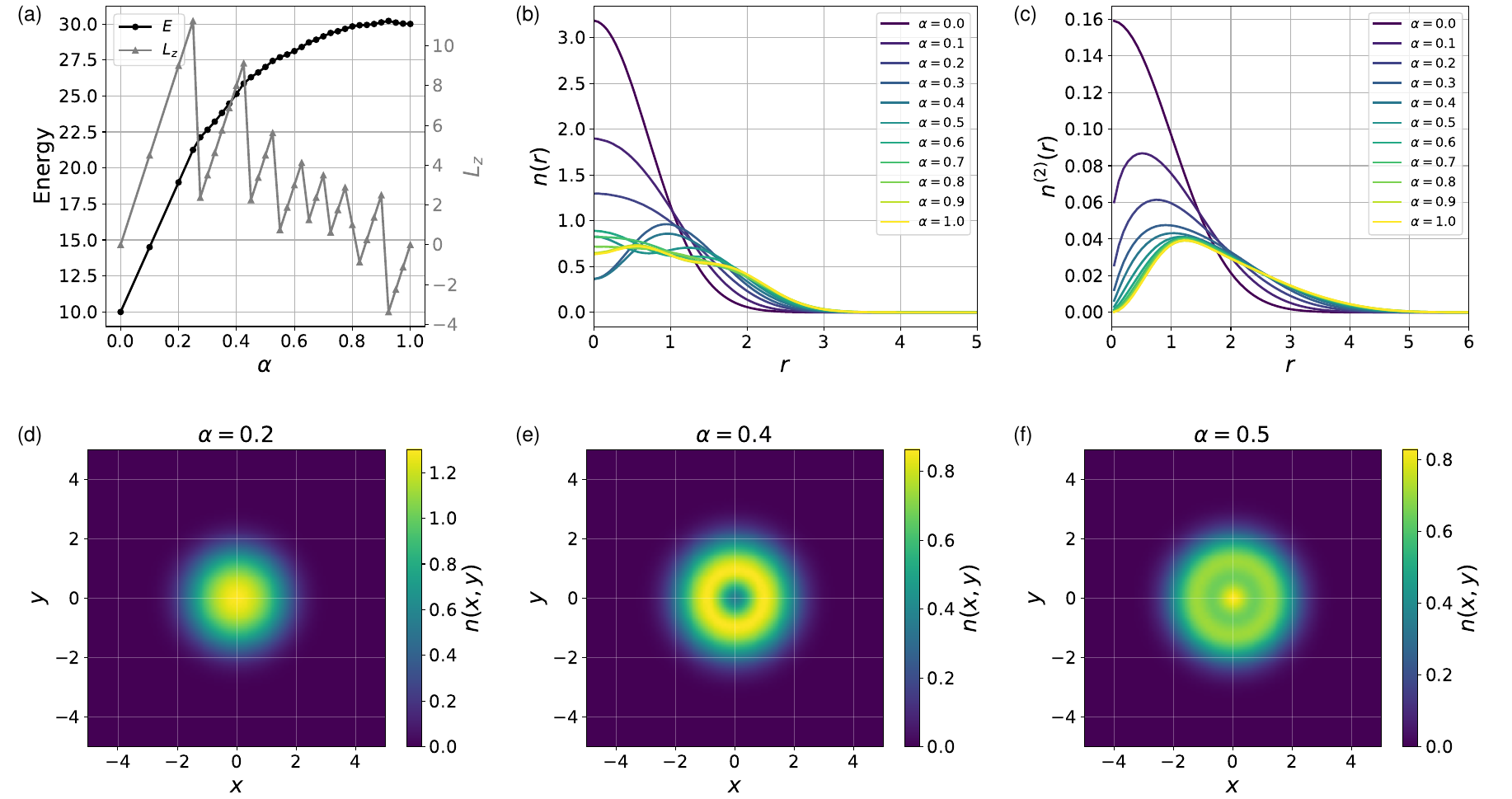}
    \caption{Results for $N = 10$ anyons. 
    Panel $(a)$ shows the ground state energy and the total angular momentum as a function of the exchange parameter $\alpha$.
    Error bars are not plotted as they are too small to be seen.
    $(b)$: Radial density $n(r)$ of the ground state wavefunction. At $\alpha \approx 0.4$, the density develops a hole in the center of the harmonic well. 
    $(c)$: Translationally averaged pair-density function $n^{(2)}(r)$ for a range of $\alpha$. 
    Panels $(d)$, $(e)$, and $(f)$ correspond to the real space density at $\alpha = 0.2$, $0.4$, and $0.5$, respectively.}
    \label{fig:n10}
\end{figure*}

The ground state energy as a function of $\alpha$ for $N=10$ is shown in Figure \ref{fig:n10}$(a)$, and for $N = 6$ in Appendix \ref{chap:results_n6}.
Our calculation shows that several distinct (non-adiabatically connected) branches become the ground state at different values of $\alpha$.
Between $\alpha = 0$ and $\alpha \approx 0.25$, the ground state is in the linear, holomorphic branch.
The level crossings can be identified by looking at the total angular momentum $L_{z} = \sum_{i}[\bm{r}_{i}\times(-i\bm{\nabla}_{i} + \bm{a}_{i})]_{z}$ eigenvalue, which jumps discontinuously at the transitions.
Additional ground state energies can be found in the supplementary information, and in Appendix \ref{chap:additional_data}.

\begin{figure}[h!]
    \centering
    \includegraphics[width=\columnwidth]{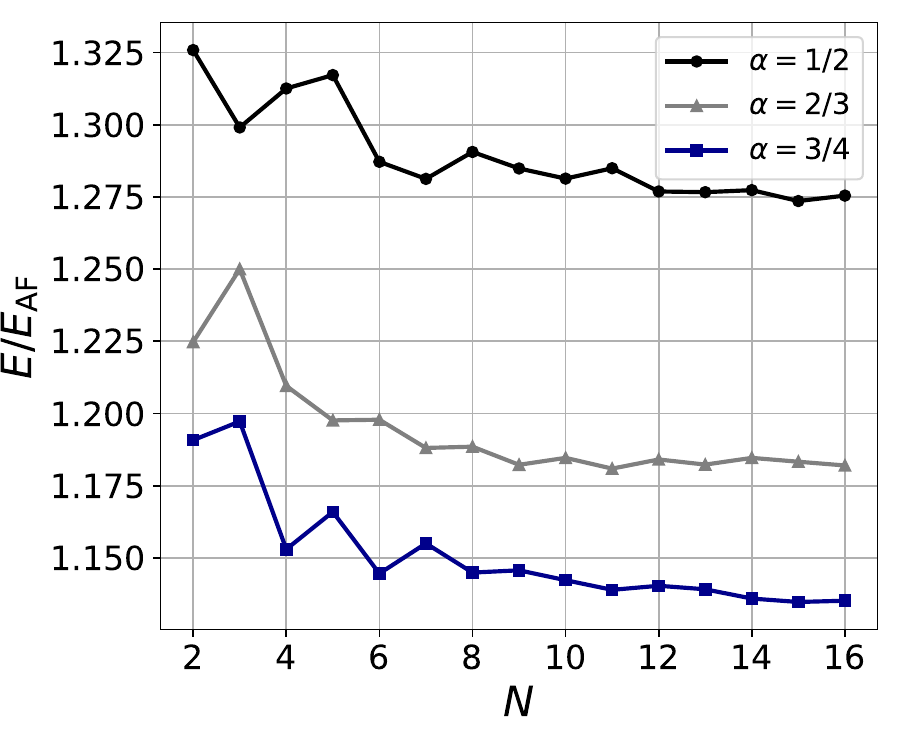}
    \caption{Scaling of ground state energy as a function of $N$ and fixed $\alpha$, as a ratio to the prediction of the average field (AF) approximation in Ref.~\cite{lundholm2015average, lundholm2017, Lundholm_review}, with $E_{\text{AF}}(N, \alpha) = \frac{\sqrt{8}}{3}\sqrt{\alpha}N^{3/2}$.}
    \label{fig:semiclassical_pred}
\end{figure}

We also investigate the scaling of the ground state energy with $N$ and $\alpha$. 
``Average field" (AF) approximations predict an energy for large $N$ of $E_{\text{AF}}(N, \alpha) = \frac{\sqrt{8}}{3}\sqrt{\alpha}N^{3/2}$ \cite{lundholm2015average, lundholm2017, Lundholm_review}. For $\alpha=1/2$, $2/3$, and $3/4$, we study system sizes up to $N=16$.
The ratio between our calculations and $E_{\text{AF}}$ is shown in Figure \ref{fig:semiclassical_pred}. 
The ratio approaches an $\alpha$-dependent constant ($\approx$1.1–1.3), so the $N^{3/2}$ scaling is reproduced, but the prefactor is only qualitatively correct.

\subsection{Density}
\label{chap:observables}
We investigate two key properties of the ground state as a function of $\alpha$: the radial density $n(r)$, and the translationally averaged pair-density $n^{(2)}(r)$, defined as
\begin{align}
    n(r) &= \frac{1}{2\pi r}\sum_{i}\langle\delta(|\bm{r}_{i}| - r)\rangle,\\
    n^{(2)}(r) &= \frac{1}{2\pi r}\frac{1}{N(N-1)}\sum_{i\ne j}\langle\delta(|\bm{r}_{i} - \bm{r}_{j}| - r)\rangle.
\end{align}
While non-interacting fermions form rings in a harmonic trap, bosons do not.
The radial density of anyons is, therefore, a non-trivial observable. 
While one might expect $n(r)$ to smoothly interpolate between the bosonic and the fermionic radial profiles, unexpected behavior can occur. 
For instance, at $N = 10$, both bosons and fermions have a high amplitude at the origin, with fermions also forming a second shell at $r\sim 1.7$.
However, near $\alpha = 0.4$, the pattern changes, and the origin is significantly depleted, as demonstrated in Figure \ref{fig:n10}($e$). 
The range of values of $\alpha$ for which this depletion is observed coincides with the region between two level crossings, suggesting that these states are connected to excited states at generic $\alpha$, but happen to become the ground state for a narrow range of values of $\alpha$.
We stress that while this is only shown for $N = 10$ (and $N = 6$ in Appendix \ref{chap:results_n6}), we have observed a similar behavior at other particle numbers, also near $\alpha \approx 0.4$.
Our calculations also demonstrate that the level crossings become very dense in $\alpha$ as the number of particles is increased.
Therefore, to see every possible ground state branch, one has to perform calculations in a very fine grid of values of $\alpha$.
For the 10 anyon calculation we needed steps in $\alpha$ of 0.025 to clearly see the level crossings.
This makes it expensive to study the whole range of $\alpha$ for larger $N$, although we are confident that if attempted, we could obtain accurate results for larger $N$.

The pair-density does smoothly interpolate between the bosonic and the fermionic distribution.
At $\alpha = 0$, the pair correlation function approaches a finite value as $r\to 0$, as expected for a bosonic system.
At any finite $\alpha$, the pair density approaches 0 instead, the correct behavior for the hard-core anyon condition.
As $\alpha$ is increased, the Fermi hole enlarges, and is maximal at the fermion ($\alpha = 1$) end.

\subsection{Comparison with the almost-fermionic perturbation theory}\label{sec: fermionic perturbation theory}
Previously we have seen that near the bosonic end the many-anyon ground state is captured by the holomorphic branch with energy $E^{(0)}_N$ in Eq.~\eqref{eq: holomorphic energy}. This can be well understood from a perturbative analysis based on bosons weakly coupled to a Chern-Simons gauge field with coupling strength $\abs{\alpha} \ll 1$ \cite{chou1991,nishida2007}, and is also seen to be consistent with our NNVMC calculations. In this section, we turn to the fermionic end, compute the 
first-order energy correction to the ground state of $N$ anyons in a harmonic trap, for $\abs{\alpha-1}\ll 1$. Our analysis here provides additional theoretical benchmarks for our NNVMC results using AnyNet for $N$ up to 20.

\begin{figure}[t]
    \centering
    \includegraphics[width=\columnwidth]{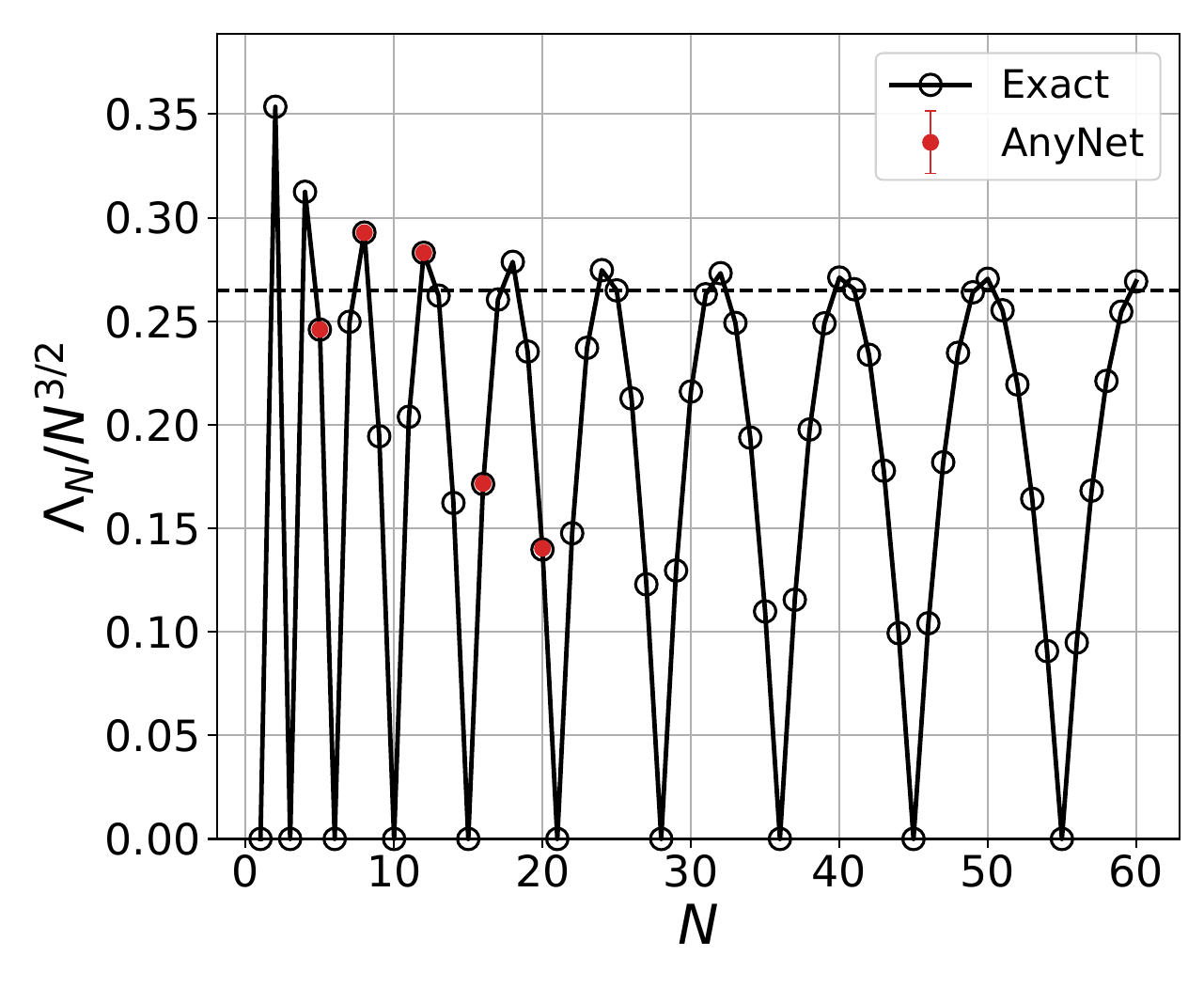}
    \caption{The linear coefficient $\Lambda_N$, characterizing the leading order statistical effect near the fermionic limit, is plotted as a function of the number of anyons $N$ from $N=1$ to 60. For selected values of $N$, the NNVMC results based on AnyNet are also shown (red solid circles). Uncertainties for the NNVMC results are too small to be displayed, and we refer to Table~\ref{tab:lambda-nn-vmc} for their values. The vertical axis is normalized as $\Lambda_N/N^{3/2}$,  which reveals an overall convergence of trend for large $N$. The maximal values of $\Lambda_N/N^{3/2}$ between two consecutive closed shell cases extrapolate to around $0.265$ (dashed line) in the infinite-$N$ limit.}
    \label{fig: lambda_N}
\end{figure}

Specifically, we are interested in the linear coefficient $\Lambda_N$ defined as follows:
\begin{equation}
    E_N(\alpha)
 =E_N(1)-\Lambda_N\abs{\alpha-1}+\mathcal{O}((\alpha-1)^2),
\end{equation}
where $E_N(\alpha)$ is the harmonic-trap ground state energy for $N$ anyons with statistics parameter $\alpha$. It is easy to see that $E_N(1) \sim \frac{\sqrt{8}}{3}N^{3/2}$ for large $N$ at the fermionic end.
Based on the fermionic-gauge Hamiltonian in Eq.~\eqref{eq:anyon-hamiltonian} and using the degenerate first-order perturbation theory, we obtain $\Lambda_N$ as follows:
\begin{equation}\label{eq: Lambda_N from M}
    \Lambda_N=\lambda_{\max}(\mathcal{M})=\max_L\lambda_{\max}\bigl(\mathcal{M}^{(L)}\bigr),
\end{equation}
where $\lambda_{\max}$ selects the largest eigenvalue of a matrix, and $\mathcal{M}$ consists of many-body matrix elements for the term in Eq.~\eqref{eq:anyon-hamiltonian} that is linear in $(\alpha-1)$, namely,
\begin{equation}\label{eq: M_PQ definition}
    \mathcal{M}_{ab} = {\binom{N}{2}} \bra{\Psi_a}\hat{V}_{12}\ket{\Psi_b},\;\;\hat{V}_{12} = \frac{\hat{\ell}_{12}}{\rho_{12}^2}.
\end{equation}
Here $\ket{\Psi_{a,b}}$ represents one of the degenerate $N$-fermion ground states, and $\hat{V}_{ij}$ is the two-body statistical interaction term in Eq.~\eqref{eq:anyon-hamiltonian} acting on particle $i$ and $j$, and we have introduced the relative coordinate $\boldsymbol\rho_{ij} = (\mathbf{r}_i - \mathbf{r}_j)/\sqrt{2}$ and the associated relative angular momentum operator $\hat{\ell}_{ij}$. Using the rotational symmetry of the problem, $\mathcal{M}=\bigoplus_L \mathcal{M}^{(L)}$ is further block diagonalized into different total angular momentum ($L$) sectors (i.e., $\mathcal{M}_{ab}=0$ if $L_a \neq L_b$). In Appendix \ref{chap: perturbation appendix}, we derive a compact analytical formula for $\mathcal{M}$, through which the exact computation of $\Lambda_N$ becomes an efficient task of computing eigenvalues of several small matrices.

Notice that the two-body operator $\hat{V}_{12}$ is odd under time-reversal, hence for specific number of particles $N$ (e.g. $N=1,3,6,10,...)$ where the free-fermion ground state is non-degenerate, we simply have $\Lambda_N=0$. 
Away from these specific numbers, we find $\Lambda_N>0$ in general as shown in Fig.~\ref{fig: lambda_N}, and thus the anyonic behavior is perturbatively attractive as compared to the fermionic case. 
In Table~\ref{tab:lambda-nn-vmc}, for a selected set of $N$ we have documented a comparison between the exact perturbation theory values with our quadratic fitting values obtained from the AnyNet NNVMC simulations with $\alpha=0.98, 0.99, 0.995, 0.9975$ and $1$ (see also the data documented in Appendix~\ref{chap:additional_data}). 
The observed close numerical match, which is also shown in Fig.~\ref{fig: lambda_N}, demonstrates the effectiveness of our AnyNet ansatze for simulating intermediate scale anyonic systems. 

\begin{table}[t]
\centering
\caption{Comparison of $\Lambda_N$ from exact first-order
perturbation theory and NNVMC fits based on the AnyNet trained ground-state energies at statistics parameters $\alpha=0.98,0.99,0.995,0.9975$ and $1$. Exact values are rounded
to at most three decimal places. Parentheses give
one-standard-error fitting uncertainties in the last quoted
digit.}
\label{tab:lambda-nn-vmc}

\footnotesize
\setlength{\tabcolsep}{2pt}
\renewcommand{\arraystretch}{1.2}

\begin{tabular*}{\columnwidth}
    {@{\extracolsep{\fill}}lccccc@{}}
\toprule
$N$ & 5 & 8 & 12 & 16 & 20 \\
\midrule
Exact
& 2.75\phantom{0(0)} & 6.625\phantom{(0)} & 11.766\phantom{(0)} & 10.969\phantom{(0)} & 12.5\phantom{0(0)} \\
AnyNet
& $2.750(2)$ & $6.624(4)$ & $11.767(1)$
& $10.986(7)$ & $12.55(1)$ \\
\bottomrule
\end{tabular*}
\end{table}

\begin{figure*}[t!]
    \centering
    \includegraphics[width=0.8\textwidth]{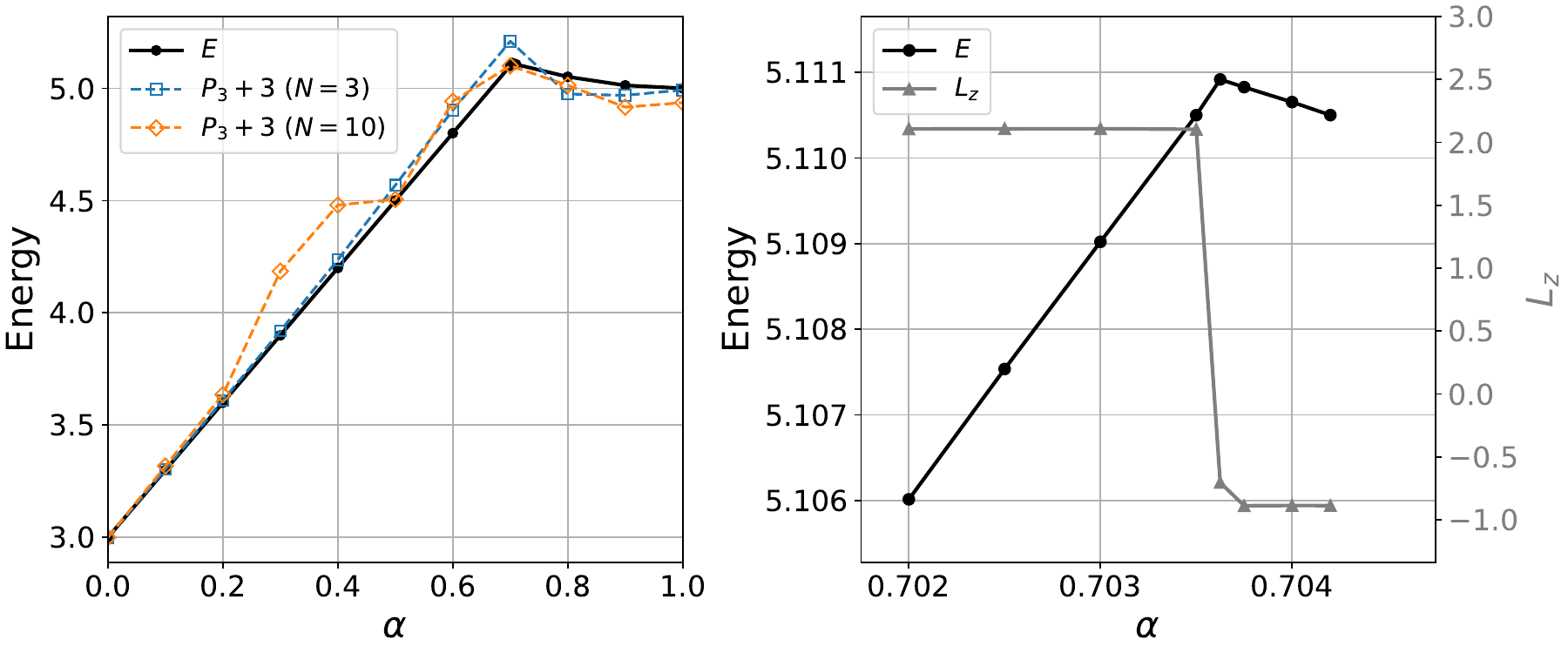}
    \caption{Left: Ground state energy $E$ for $N = 3$ anyons, together with the triple-coalescence exponent $P_3$, obtained from an $N = 3$ and an $N = 10$ calculation.
    If $M$ anyons are a distance $\rho$ from each other, then $P_M$ is defined by the scaling $|\psi| \sim \rho^{P_M}$. $M=3$ was chosen here as the smallest non-trivial example.
    We find good agreement between the extracted $P_3$ and the NRCFT prediction in Eq.~\eqref{eq:nrcft-clustering-power}.
    Right: Energy for $N = 3$ near the transition between the linear holomorphic branch and the non-linear branch, which we conclude sits between $\alpha = 0.7035$ and $0.70375$.
    Between these two values, the angular momentum eigenvalue changes discontinuously, indicating a level crossing.}
    \label{fig:coalescence}
\end{figure*}

\subsection{Wavefunction scaling at coalescence}
\label{chap:coalescence_results}
As derived in Section \ref{chap:coalesecence}, the scaling of the anyonic wavefunction $\psi$ near a coalescence of $M$ anyons is related to the energy of $M$ anyons at the same $\alpha$ in a harmonic trap, see Eq.~\eqref{eq:nrcft-clustering-power}.
We define the radius of a cluster $C$ of $M$ anyons as 
\begin{align}
    \rho_{C}^{2} &= \frac{1}{M}\sum_{i\in C}|\bm{r}_{i} - \bm{R}_{C}|^{2},\\
    \bm{R}_{C} &= \frac{1}{M}\sum_{i\in C}\bm{r}_{i}.
\end{align}
The coincidence exponent $P$ can then be defined as $|\psi| \sim \rho_{C}^{P}$ as $\rho_{C} \to 0$.
$P$ is evaluated by estimating the quantity
\begin{align}
    D(\rho_{C}) &= \rho_{C}\partial_{\rho_{C}}\log|\psi| = \sum_{i\in C}(\bm{r}_{i} - \bm{R}_{C})\cdot\bm{\nabla}_{i}\log|\psi|,
\end{align}
which obeys $D(\rho_{C} \to 0) = P$.
During inference, samples distributed by $|\psi|^{2}$ are generated.
For each sample, all the $\begin{pmatrix}
    N\\M
\end{pmatrix}$ clusters are obtained, and for every cluster, its radius $\rho_{C}$ and $D(\rho_{C})$ are calculated.
Then, the calculated $\{(\rho_{C,i}, D(\rho_{C})_{i})\}$, with $i$ ranging over all clusters sampled, are binned and averaged in $\rho_{C}$.
Let the bin-aggregated $\{D(\rho_{C})_{i}\}$ be $\bar{D}(\rho_{C})$.
$\bar{D}(\rho_{C})$ is fitted against $\rho_{C}$, to a cubic polynomial, with each data point weighted inversely proportional to the reciprocal of the variance.
Our reported estimate of $P$ is the extrapolation to $\rho_{C} = 0$ of the cubic fit.

In Figure~\ref{fig:coalescence}, we present the triple-coalescence ($M=3$) analysis using both $N=3$ and $N=10$ AnyNet trained states at various $\alpha$. The extracted coalescence exponents $P_3$ at different $\alpha$ are found to be in agreement with the NRCFT prediction obtained from Eq.~\eqref{eq:nrcft-clustering-power}. We stress that a good agreement with the theoretical prediction is not a trivial result.
While the exact behavior among pairs is enforced, no restrictions are enforced for larger clusters.
As the wavefunction is also very small in the region near coalescence of multiple anyons, the probability of drawing random samples when multiple anyons are close is suppressed, meaning that the network had little data to train in this region of configuration space.
Further, only the smallest, leading power is related to the ground state energy of $M$ anyons in a harmonic trap; there are also higher-order terms related to excited states.
As we can only sample points at a finite $\rho_{C}$, our data is also polluted with these higher-order contributions, making the extraction of the leading power difficult in practice.
\begin{figure}[h!]
    \centering
    \includegraphics[width=\columnwidth]{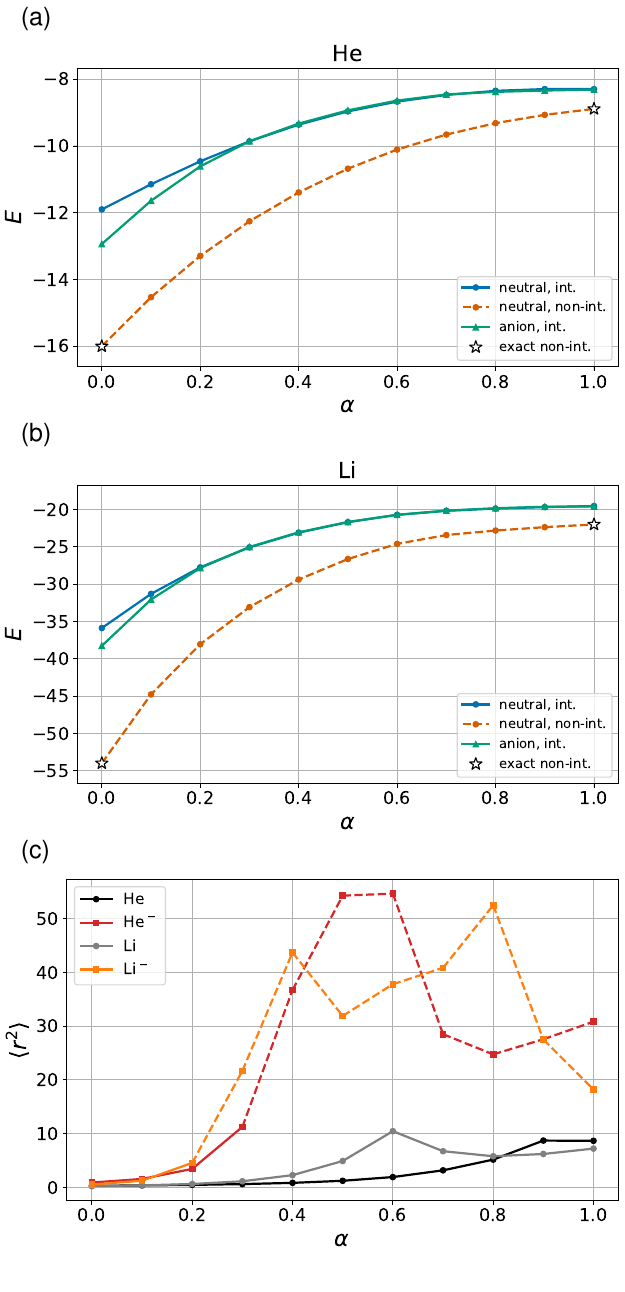}
    \caption{($a$): Ground state energy of helium ($Z = 2$) anyonic atoms and anions (negatively charged ions), interacting and non-interacting.
    The exactly known ground state energies are marked with a star.
    ($b$): Similar to ($a$), but for lithium ($Z = 3$).
    ($c$): $\langle r^{2}\rangle$ for the ground state of interacting anyonic He, He$^{-}$, Li and Li$^{-}$.
    The neutral atoms have a low $r^{2}$ expectation, suggesting that they are bound for all $\alpha$.
    Anyonic Li is least localized at an intermediate $\alpha$ of around 0.6.
    However, the anions only have a small $\langle r^{2}\rangle$ for small $\alpha$, suggesting that an additional anyon may bind the atom.
    For larger $\alpha$, the anyon cloud becomes very large, which we interpret as the additional anyon not binding to the atom. 
    These data points are marked with dashed lines.
    }
    \label{fig:atoms_r2}
\end{figure}
\subsection{Anyonic atoms}
\label{chap:anyonic_atoms}
Finally, we have investigated the behavior of four anyon atomic systems, He, He$^{-}$, Li, and Li$^{-}$, where the element refers to the charge of the nucleus ($Z_{\text{He}}=2, Z_{\text{Li}}=3$), and the superscript refers to whether an extra anyon has been added.
All atomic calculations are with anyons interacting via the Coulomb potential unless otherwise stated.
We note that the short-range behavior of Coulomb-interacting anyons is modified at a subleading order, as discussed at the end of Appendix \ref{chap:short_range}. This effect has been taken into account by adopting modified short-range channel functions in AnyNet as discussed below Eq.~\eqref{eq: chi_c}.

The ground state energy as a function of the statistics parameter is shown in Figure \ref{fig:atoms_r2}, in panel ($a$) for He and panel ($b$) for Li.
The ground state energy is exactly known (shown with a star) for the non-interacting atoms at $\alpha=0$ and $\alpha = 1$, which match our calculations to a few micro-Hartree.
At intermediate $\alpha$, the energy increases monotonically, and the wavefunctions consistently converged to a valid angular momentum.
For interacting atoms, at small $\alpha$, the anion has a significantly lower energy than the neutral atom, indicating the anion is stable.
Theoretically, one can work out if an anion is stable by computing $B(X) = E(X) - E(X^{-})$, where $E(X)$ is the ground state energy of system $X$.
If $B(X)$ is positive, then the additional anyon binds to the atom.
If the additional anyon does not bind to the atom, then the ground state energy of $X^{-}$ contains the additional anyon delocalized infinitely far away with 0 energy, so $E(X^{-})=E(X)$, and $B(X) = 0$.
However, this is difficult to do in practice, as we require that the wavefunction vanishes far from the atom, which is enforced by applying an exponential envelope to the ansatz.
For a system that does not bind, the bulk of the network becomes large to overcome this exponential suppression as much as possible.
This requires the size of the parameters to become large, which causes training to be very slow and eventually unstable.

A more controlled approach for studying anyonic anion is to train for a fixed number of iterations and to calculate $\langle r^{2}\rangle$ as a measure of how delocalized the anyon cloud has become.
This is shown in Figure \ref{fig:atoms_r2}, panel ($c$).
The neutral atoms have a small $\langle r^{2}\rangle$ for all $\alpha$, indicating that the neutral atoms are bound.
The anions only have a small $\langle r^{2}\rangle$ for $\alpha$ lower than around 0.2-0.3, matching the conclusion from the energy eigenvalues.
Interestingly, neutral anyonic Li is least localized at an intermediate $\alpha$ of around 0.6, where there is a level crossing.
Anyonic He also has a level crossing, near $\alpha = 0.9$, which is not present in the non-interacting harmonic trap with $N = 2$ (or the non-interacting He atom).
The anions have a small $\langle r^{2}\rangle$ only for small $\alpha$.
For larger $\alpha$ the anyon cloud grows and keeps increasing in size as the number of iterations is increased, which we interpret as the anion being unstable.
Note that the precise value of any observables such as the energy, angular momentum, or $\langle r^{2}\rangle$ for anyonic anions where the estimated $\langle r^{2}\rangle$ is large are not to be trusted.
These are shown only for completeness, but our exponentially confined wavefunctions cannot represent a state with a particle delocalized at infinity.
Instead, these data points (shown with dashed lines in Fig.~\ref{fig:atoms_r2}(c)) should only be interpreted as evidence for the additional anyon being delocalized.
The data is available in the supplementary information and Appendix \ref{chap:additional_data}.

\begin{figure}[h!]
    \centering
    \includegraphics[width=\columnwidth]{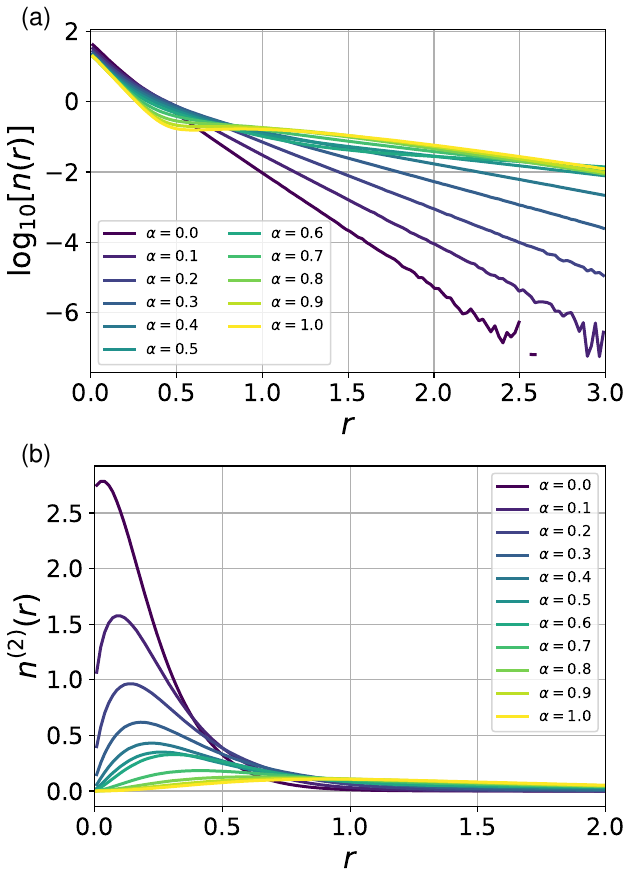}
    \caption{Densities of the interacting anyonic lithium atom. 
    Panel ($a$) shows the radial density, and panel ($b$) shows the pair density.
    The noise near the end of the radial density is due to a lack of random samples far from the nucleus.
    }
    \label{fig:atoms_density}
\end{figure}

The radial and pair densities of the interacting lithium atom ($Z = 3$, $N = 3$) as a function of $\alpha$ are shown in Figure \ref{fig:atoms_density}.
In the bosonic regime, a clear exponential decay is observed in the radial density.
With increasing $\alpha$, the state initially decays exponentially, develops a small plateau at around $r\sim 0.7$, and continues to decay exponentially but with a smaller decay rate.
The pair-density is also reminiscent of the behavior found in the harmonic trap, slightly modified due to the Coulomb interactions.
This is most easily seen in the $\alpha = 0$ pair-density, which instead of flattening at $r = 0$, requires a positive gradient at the origin.
This is further verification that our method for enforcing the correct short-range behaviour works as intended.

\section{Conclusion and Future Directions}
\label{chap:conclusion}
In this work, we present the AnyNet architecture, an anyonic neural-network ansatz for large-scale simulations of many-anyon systems.
We develop a method for analytically enforcing a desired short-range behavior amongst anyons through modified neural-Slater determinants, which could have applications in other correlated quantum systems.
We use the AnyNet to study Abelian anyons in a harmonic trap and a Coulomb trap (anyonic atoms).

In the harmonic trap, we compare our results to exact diagonalization for $N = 4$ anyons, the highest that has been done to date (in the continuum and the infinite plane).
We find excellent agreement, a few parts in a million for all exchange parameters $\alpha$.
We then study a larger system of $N = 10$ anyons, in which a very fine grid of $\alpha$ had to be used due to the rapidly increasing number of level crossings. 
The level crossings were identified by computing the total angular momentum eigenvalue, which is observed to sharply jump at transitions.
The radial density and pair density of these states were also investigated.
To benchmark our numerical results for larger systems, we compare our near-fermion results to a perturbative analysis.
At the largest system studied ($N = 20$), we agree with the theoretical prediction within 0.4\%. 
Far away from the fermionic limit, at $\alpha=1/2$, $2/3$, and $3/4$, we study system sizes up to $N=16$.
Next, we test the non-relativistic conformal field theory prediction for the scaling of the wavefunction as multiple anyons coalesce. 
The leading power of this scaling when $M$ anyons participate is predicted to be $E_M - M$, where $E_M$ is the ground-state energy of $M$ anyons in a harmonic trap.
We find good numerical agreement for $M = 3$ from an $N = 3$ and an $N = 10$ AnyNet calculation.
Lastly, we studied anyonic atoms (both non-interacting, and with a Coulomb interaction) as a toy model for anyon binding to impurities, and as a demonstration of how our method generalizes to other systems.
We study the stability of adding an extra anyon as a function of $\alpha$, and find stability only for small $\alpha$ ($\lesssim 0.2-0.3$).
We also compare to exact energy benchmarks at the fermionic and the bosonic end, agreeing to a few parts in a million.

This work constitutes a first step in large-scale simulations of intrinsically anyonic systems. 
A natural next step is to address directly the long-standing question of anyon superfluidity~\cite{laughlin1988, fetter1991, hanna1989, chen1989}. 
Several signatures of superfluidity, such as an even-odd effect in the ground state energy, a non-classical moment of inertia~\cite{leggett1970can}, and the scissors mode in an anisotropic trap~\cite{guery1999scissors}, are accessible within the harmonic trap, the latter becoming feasible with recent developments in excited-state variational Monte Carlo~\cite{pfau2024excited, wheeler2024ensemble}.

Nevertheless, the confining potential of the harmonic trap makes it challenging to confidently detect signatures of superfluidity in these systems. 
Generalizing these methods to compact geometries, such as a torus or a sphere, is a promising step towards clearly demonstrating the presence or absence of superfluidity in these systems. 
These geometries bring their own subtleties: on the torus, anyons have degenerate, multi-component wavefunctions~\cite{einarsson1990fractional}, while on the sphere the allowed values of $\alpha$ are constrained by the particle number.

Superfluidity is, however, not the only proposed fate of the anyon gas. It was recently argued that the statistical interaction alone may drive the system into a crystalline state~\cite{komargodski2026anyoncrystallizationstatistics}. 
Since our ansatz makes no assumptions about the structure of the ground state, it is well suited to distinguish between these scenarios.

Another natural extension is to include a uniform magnetic field. At a weak field, our method can be used to look for a Meissner-type effect as a possible signature for anyon superconductivity \cite{chen1989, dft_anyon_Jain2021}. 
At a sufficiently strong field, the anyons may themselves form incompressible quantum Hall states, as in the hierarchy construction~\cite{halperin1984}. 
Studying the crossover between the free anyon gas and this quantum Hall regime would connect the intrinsically anyonic Hamiltonians studied here to the fractional quantum Hall setting in which anyons are usually encountered, as well as to recent proposals for doped fractional quantum anomalous Hall insulators~\cite{shi2024, shi2025, shi2025a, mehta2026, shi2026, shi2026a, senthil2026, han2026a, Wang_Zhang_ASF_2026, zhang2026, zhang_holon_2026, lotric2026, fan2026hiddenweakpairingsuperconductivitynoninteracting, wang2026topologicalsuperconductivityabelianfractional}.

Finally, the methods presented here can in principle be generalized to non-Abelian anyons. 
In this case, the wavefunction becomes a vector in a degenerate fusion space, and exchanging two particles acts on it as a matrix, given by a representation of the braid group, rather than as a phase. 
The singular gauge transformation of Eq.~\eqref{eq:singular-gauge-transformation} then generalizes to a non-Abelian Chern-Simons gauge field, and the short-distance behavior of Appendix~\ref{chap:short_range} becomes dependent on the fusion channel of the coalescing anyons. 
As in the Abelian case, the non-relativistic conformal symmetry provides exact and perturbative results for non-Abelian anyons in a harmonic trap~\cite{doroud2018conformal}, which would serve as valuable benchmarks. 
The main challenge is that the dimension of the fusion space grows exponentially with the number of anyons.

\section*{Acknowledgments}
We thank Matthew Foulkes, Luca Delacrétaz, Hart Goldman, Guangyue Ji, Qianrui Liu, Umang Mehta, Max Metlitski, David Pfau, Shinsei Ryu, Dam Thanh Son, James Spencer, Taige Wang, Xiaochuan Wu, and Siwei Zhong for helpful discussions.
The authors thank the Gauss Centre for Supercomputing e.V. (GCS) for providing computing time on the Supercomputer JUPITER at the Jülich Supercomputing Centre via the John von Neumann Institute for Computing. JUPITER is supported by the EuroHPC JU and GCS through funding by the European Commission, the German Federal Ministry of Research, Technology and Space, and the Ministry of Culture and Science of the State of North Rhine-Westphalia.
The authors acknowledge the use of resources provided by the Isambard-AI National AI Research Resource (AIRR)~\cite{mcintosh2024isambard}.
Isambard-AI is operated by the University of Bristol and is funded by the UK Government’s Department for Science, Innovation and Technology (DSIT) via UK Research and Innovation; and the Science and Technology Facilities Council [ST/AIRR/I-A-I/1023]. We also acknowledge the use of Claude Code by Anthropic and Codex by OpenAI for assisting on derivations and code implementations.
APF's PhD work is supported by the UK Engineering and Physical Sciences Research Council under grant EP/W524323/1.
PMT is supported by postdoctoral research fellowships at the Princeton Center for Theoretical Science and the Leinweber Institute for Theoretical Physics at the University of Chicago.
LS is supported by the MIT Pappalardo Fellowship.

\newpage

\bibliography{leyna_references.bib, ref.bib}

\clearpage
\onecolumngrid

\appendix

\section{Short-Range Behaviour}
\label{chap:short_range}
In this section, we derive the short-range behaviors of Abelian anyons, which we make heavy use of throughout the manuscript. We first consider the short-range behavior for non-interacting (``free") anyons, and we note that this has been discussed previously in Ref. \cite{arovas1985}.
The calculation is easiest to do in the anyon gauge, but we will translate the result to the fermion gauge at the end.
We consider the Hamiltonian for two free anyons, and study the form of the wavefunction near coincidence.
\begin{align}
    H_\text{free} = -\frac{1}{2}\nabla^{2}_{1} -\frac{1}{2}\nabla^{2}_{2} = -\frac{1}{4}\nabla^{2}_{\bR} - \nabla^{2}_{\br},
\end{align}
where we have transformed the Hamiltonian into center of mass ($\bR$) and relative ($\br$) coordinates. The center of mass component plays no role in this analysis, so we inspect
\begin{align}
    H_{\rm rel} = -\nabla^{2}_{\br} = -\left[\frac{\partial^{2}}{\partial r^{2}} + \frac{1}{r}\frac{\partial}{\partial r} + \frac{1}{r^{2}}\frac{\partial^{2}}{\partial \theta^{2}}\right].
\end{align}
As we are dealing with hard-core anyons, $r = 0$ is a regular singular point, where the wavefunction vanishes. Near $r=0$, we can expand the anyonic wavefunction (in the relative sector) as
\begin{align}
    \psi_A(\br) = f_n(r) e^{i(\alpha + 2n)\theta},\\
    f_n(r) = r^{\gamma_n}\sum_{m=0}a_{m}r^{m},
\end{align}
where $\gamma_n\geq 0$. The choice of the relative angular momentum set to $\alpha + 2n$, $n \in \mathbb{Z}$, corresponds to the anyon exchange condition.
Without loss of generality, we will insist on $a_{0} \ne 0$, corresponding to $\gamma_n$ being the minimal non-trivial exponent in this channel.
Inserting this ansatz into the Schr\"odinger equation, we obtain
\begin{align}
    -&\sum_{m=0}(\gamma_n + m)(\gamma_n + m - 1)a_{m}r^{\gamma_n + m - 2} 
    -\sum_{m=0}(\gamma_n + m)a_{m}r^{\gamma_n + m - 2} +\sum_{m=0}(\alpha + 2n)^2a_{m}r^{\gamma_n + m - 2}
    = E_{\rm rel}\sum_{m=0}a_{m}r^{\gamma_n + m}.
\end{align}
Matching the coefficients on both sides, we obtain
\begin{align}
    \gamma_n^2 = (\alpha + 2n)^2 \implies \gamma_n = |\alpha + 2n|,
\end{align}
where we have rejected the negative $\gamma$ solutions based on our choice to focus on hardcore anyons.
From this, we conclude that the allowed short-distance behaviors for $\psi_A(r)$ are
\begin{equation}
    \psi_A \sim r^{|\alpha+2n|} e^{i(\alpha+2n)\theta},\;\;(n\in\mathbb{Z})\label{eq:short_range_anyon_gauge}
\end{equation}
up to multiplication of a smooth function that can be easily represented by a neural network. Translating this result into the fermion gauge, using Eq.~\eqref{eq:singular-gauge-transformation}, we obtain
\begin{equation}
    \psi_F \sim r^{|\alpha+2n|} e^{i(1+2n)\theta},\;\;(n\in\mathbb{Z})
\end{equation}
which is the short-distance behavior quoted in Eq.~\eqref{eq:channels}.

Next, we consider how the Coulomb interaction between anyons modify the short-distance behaviors, which is incorporated into the AnyNet for studying the interacting anyonic atom in Sec. \ref{sec: NNVMC results}. To that end, the relative sector $H_{\rm rel}$ contains an extra $1/r$ term (recall we adopt the Hartree units, see Eq.~\eqref{eq: anyonic atom Hamiltonian} in the main text), and repeating the above analysis by substituting the above expansion ansatz into the Schr\"odinger equation, upon matching coefficients we obtain
\begin{equation}
    \gamma_n = \abs{\alpha+2n} \quad \text{and} \quad a_1 = \frac{a_0}{2\gamma_n+1}. 
\end{equation}
Hence, statistics still fixes the leading power in the same way, while the Coulomb interaction fixes the next coefficient. In the fermion gauge, the short-distance behavior of the Coulomb-interacting anyonic wavefunction thus takes the form
\begin{equation}
    \psi_F \sim r^{|\alpha+2n|} e^{i(1+2n)\theta} \left[1+\frac{r}{2|\alpha+2n|+1}\right].
\end{equation}
This is the kind of short-range behavior that we build into the AnyNet for studying the interacting anyonic atom. See also the discussion in the main text around Eq.~\eqref{eq: chi_c}. 

\section{Non-Relativistic Conformal Symmetry of the Free Anyon Hamiltonian}
\label{chap:NRCFT_app}
In this section, we follow the first-quantized analysis of~\cite{werner2006}, which was originally applied to the Fermi gas at unitary, to expose the structure of the free anyon eigenstates as a consequence of a non-relativistic conformal symmetry. 

To keep things self-contained, we restate our model here, and also make re-introduce the constants set to $1$ in the main text, in order to keep units explicit. The Hamiltonian is
\begin{equation}
H=
\frac{\hbar^2}{2m}\sum_{i=1}^N
\left[-i\nabla_i+\mathbf a_i\right]^2
+\frac{m\omega^2}{2}\sum_{i=1}^N |\mathbf r_i|^2 ,
\label{eq:anyon-hamiltonian-app}
\end{equation}
where
\begin{equation}
\mathbf a_i
=
(\alpha-1)\sum_{j\neq i}
\frac{\hat{\mathbf z}\times(\mathbf r_i-\mathbf r_j)}
{|\mathbf r_i-\mathbf r_j|^2}.
\label{eq:statistical-gauge-field-app}
\end{equation}
We define the operators
\begin{equation}
    \begin{aligned}
        \bm{R} &\equiv (\bm{r}_1, \bm{r}_2, \ldots, \bm{r}_N)\,,
        \\
        R^2 &\equiv \sum_i \abs{r_i}^2\,,
        \\
        \hat{D} &\equiv \bm{R} \cdot \partial_{\bm{R}}\,,
        \\
        L_{\pm} &= \pm  N \pm \hat{D} + \frac{H}{\hbar \omega} - m\omega R^2/\hbar\,.
    \end{aligned}
\end{equation}
One can verify
\begin{equation}
\begin{aligned}
    \comm{H}{L_{\pm}} &= \pm 2 \hbar \omega L_{\pm}\,,
    \\
    \comm{L_+}{L_-} &= - \frac{4 H}{\hbar\omega}\,.
    \end{aligned}
\end{equation}
We define the Casimir operator
\begin{equation}
    \begin{aligned}
        \hat{\mathcal{C}} &\equiv H^2 - \frac{1}{2} (\hbar\omega)^2 \left( L_+ L_- + L_- L_+\right)
    \end{aligned}
\end{equation}
which commutes with $H$ and $L_{\pm}$. For a ladder of energy eigenstates with constant $\hat{\mathcal{C}}$, the Casimir takes the value $C = E_g (E_g -2 \hbar\omega)$, where $E_g$ is the energy of the lowest-energy state of the ladder (annihilated by $L_-$). This implies that $E_g$ can be obtained by the operator
\begin{equation}
    H_g \equiv \hbar \omega + \sqrt{\hat{\mathcal{C}} + (\hbar\omega)^2}\,.
\end{equation}
We define the bosonic operators $b\,, b^\dagger$ as
\begin{equation}
    \begin{aligned}
        b &\equiv \sqrt{\frac{\hbar\omega}{2(H + H_g)}}L_-\,,
        \\
        b^\dagger &\equiv L_+ \sqrt{\frac{\hbar \omega}{2(H + H_g)}}\,.
    \end{aligned}
\end{equation}
By introducing these operators, the Hamiltonian can be written exactly as
\begin{equation}
    H = H_g + 2 \hbar\omega b^\dagger b\,.
\end{equation}
So, the spectrum of the Hamiltonian splits into sectors with energies $E_g\,, E_g + 2 \hbar \omega\,, E_g + 4 \hbar\omega\ldots$.

The lowest-energy state within a given tower, which includes the global ground state, obeys the equation $L_- \ket{\psi} = 0$. Integrating this equation yields
\begin{equation}
    \psi(\bm{R}) = e^{-R^2/2a^2} R^{E_{g}/(\hbar \omega) - N}f(\bm{n})
\end{equation}
where $\bm{n} \equiv \bm{R}/R$ is the unit vector, and $a = \sqrt{\hbar/m \omega}$ is the length scale of the harmonic oscillator . Hence, the ground state of an $N$-anyon system can be written as a Gaussian envelope $e^{-R^2/2a^2}$ times a homogenous function of degree $E_{g}/(\hbar \omega) - N$. 

This fixes the coalescence condition in the anyon wavefunction as $M$ anyons approach each other. For the ground state of $N$ anyons, the ground state wavefunction as all anyons approach each other must vanish as $r^{E_N/(\hbar \omega) - N}$, where $E_N$ is the ground state energy of $N$ anyons. When $M < N$ anyons approach each other, the coalescence behavior is specified by the spectrum of the $M$-anyon problem, with the leading-order exponent related to the $M$-anyon ground state energy, $E_M/(\hbar \omega) - M$, although subleading exponents corresponding to excited states in the $M$-anyon spectrum will also exist. The case of $M=2$ particles corresponds to the analysis in App.~\ref{chap:short_range}; the spectrum of the two anyon problem decomposes into two towers of energies, one with lowest energy $\hbar \omega (2 + \alpha)$ and the other with lowest energy $\hbar \omega (4 - \alpha)$. The two short-distance channels in our ansatz are necessary to capture both possibilities. 

\section{Divergent Monte Carlo Variance}
\label{chap:variance}
The many-body wavefunction must satisfy the short range behavior derived in Eq. \eqref{eq:short_range_anyon_gauge} for every pair of particles.
We show that failure to represent this short-range behavior results in a strong local energy ($E_{L}$) divergence, which causes the variance of the estimated energy expectation value to diverge.

We perform this calculation in the anyon gauge, and consider only the relative coordinates in complex representation, $z \equiv (x_{1} - x_{2}) + i(y_{1} - y_{2})$.
In this form, $H_{r} = -4\partial\bar{\partial}$, where $\partial$ ($\bar{\partial}$) is the holomorphic (antiholomorphic) derivative.
As we are working in the anyon gauge, we require the phase be $\exp[i\theta (\alpha + 2n)]$, with $\theta = \arg(z)$, $n\in\mathbb{Z}$, and allow for any radial power (we only consider the leading radial power).
Such state can be generally written near coincidences as $\psi_{A}(z, \bar{z}) = z^{\gamma + \alpha + 2n}\bar{z}^{\gamma}$.
\begin{align}
    E_{L} &\propto z^{-(\gamma + \alpha + 2n)}\bar{z}^{-\gamma} \partial_{z}\partial_{\bar{z}}z^{\gamma + \alpha + 2n}\bar{z}^{\gamma},\\
    &= \frac{(\gamma + \alpha + 2n)\gamma}{z\bar{z}}.
\end{align}
The coefficient of the $1/r^{2}$ vanishes if $\gamma = 0$ (one of the channels in the positive branch) or $\gamma + \alpha + 2n = 0$ (one of the channels in the negative branch).
Let $\Gamma (= 2\gamma + \alpha + 2n)$, be the leading radial power of the state. Consider now the Monte Carlo variance (ignoring the finite term $\langle E_{L}\rangle^{2}$) due to random samples near coincidence of particles,
\begin{align}
    \sigma^{2} &\sim\int_{0}^{R}|\psi|^{2} \left(\frac{1}{r^{2}}\right)^{2}rdr,\\
    &\sim \int_{0}^{R}r^{2\Gamma - 3}dr,
\end{align}
which diverges near $R = 0$ unless $\Gamma > 1$. 
However, generic anyon wavefunctions have $\Gamma \le 1$.
Therefore, any ansatz with the correct radial power $\Gamma$, will have a divergent Monte Carlo variance unless the short-range structure exactly matches the ones derived in Appendix \ref{chap:short_range}.

\section{Optimization}
\label{chap:optim}
The network parameters were optimized to minimize the expectation value of the Hamiltonian.
\begin{align}
    \langle H\rangle &= \frac{\int d\bm{R}\bar{\psi}(\bm{R})H\psi(\bm{R})}{\int d\bm{R} \bar{\psi}(\bm{R})\psi(\bm{R})} = \langle E_{L}(\bm{R})\rangle_{\bm{R}\sim |\psi(\bm{R})|^{2}},
\end{align}
where $E_{L}(\bm{R}) = \frac{1}{\psi(\bm{R})}H\psi(\bm{R})$ is the local energy, and $\bm{R}$ is shortand for $(\bm{r}_{1}, \dots, \bm{r}_{N})$.
The gradient with respect to variational parameters can be computed analytically as
\begin{align}
    \partial_{\theta_{i}}\langle H\rangle &= 2\Re[\langle E_{L}\bar{O}_{i}\rangle - \langle E_{L}\rangle\langle \bar{O}_{i}\rangle],\\
    O_{i} &= \partial_{\theta_{i}}\log(\psi),
\end{align}
which is evaluated with Monte Carlo integration.
The parameters were optimized with KFAC \cite{martens2015optimizing}, details of the training and network hyperparameters used can be found in Table \ref{table:optim}.

\begin{table}[h!]
\centering
\begin{tabular}{c c}
  \hline
  Parameter & value\\
  \hline
  Pretraining iterations & 0\\
  Dimensions & 2\\
  Complex Output & True\\
  Laplacian & folx\\
  Envelope & $\exp\left(-\frac{1}{2}\sum_{i}r_{i}^{2}\right)^{*}$\\
  Deep Set number of layers & 2\\
  Deep Set nodes per layer & 256\\
  Deep Set activation function & tanh\\
  \hline
\end{tabular}
\caption{Table of hyperparameters used. Any hyperparameter not referenced was taken to be the default value~\cite{ferminet_github}. (*): For the harmonic trap; otherwise, the default exponentially decaying envelope.\label{table:optim}
}
\end{table}

\section{Extrapolation of Exact Diagonalization Energies}
\label{chap:ED_extrapolation}

In this appendix we describe the exact diagonalization used to obtain the $N = 4$
energies quoted in the main text. The construction follows~\cite{sporre1992}, to which we refer the reader for the original derivation.

Starting with a wavefunction $\Psi$ in the bosonic gauge, we perform a similarity transformation
$\Psi = \prod_{i<j} \abs{r_i - r_j}^{\alpha}\, \Phi$. This similarity transformation removes the three-body interaction in the Hamiltonian, and the reduced
Hamiltonian acting on $\Phi$ is \textit{non-Hermitian} while retaining a real
spectrum. For $N = 4$, we can use relative coordinates
\begin{equation}
  \begin{aligned}
    x_1 &= \tfrac{1}{2}(z_1 + z_4 - z_2 - z_3)\,, \\
    x_2 &= \tfrac{1}{2}(z_2 + z_4 - z_1 - z_3)\,, \\
    x_3 &= \tfrac{1}{2}(z_3 + z_4 - z_1 - z_2)\,,
    \label{eq:moshinskyCoordinates}
  \end{aligned}
\end{equation}
which, together with $X = \tfrac{1}{2}(z_1 + z_2 + z_3 + z_4)$, constitute an
orthogonal change of variables. The oscillator part of the Hamiltonian is therefore
unchanged and the physical center of mass $X$ separates exactly. Crucially, every
pair separation is then a simple combination of two Moshinsky coordinates: one has
$z_1 - z_2 = x_1 - x_2$ and $z_4 - z_3 = x_1 + x_2$, and likewise for the remaining
pairs, so that the six separations are exhausted by $x_k - x_l$ and $x_k + x_l$ with
$k < l$.

We expand each $x_k$ in the circular oscillator basis $\ket{n_{a}, n_{b}}$, whose
states carry energy $n_{a} + n_{b} + 1$ and angular momentum $n_{a} - n_{b}$. We
write $m_k$ for the angular momentum carried by the $k$-th coordinate and
$j = m_1 + m_2 + m_3$ for the total, which is conserved and blocks the Hamiltonian.
The permutation group $S_4$ contains the subgroup $S_3 \ltimes D_2$, where $S_3$ permutes the
$x_k$ and $D_2$ reverses the signs of two of them. Performing the $D_2$ sum
explicitly restricts the basis to states in which every $m_k$ shares the parity of
$j$, and on that restricted sector the matrix elements of the $x_k + x_l$ pairs
coincide with those of the $x_k - x_l$ pairs. The former therefore merely double the
latter, and the four-anyon problem collapses onto a \textit{three}-coordinate problem
with doubled coupling,
\begin{equation}
  \begin{aligned}
    H(\alpha) = H_0(x_1, x_2, x_3) + 2\alpha W\,, \qquad
    W = -2 \sum_{k<l} \frac{\partial_{x_k} - \partial_{x_l}}
                           {\bar{x}_k - \bar{x}_l}\,,
    \label{eq:moshinskyReduced}
  \end{aligned}
\end{equation}
where $H_0$ is the oscillator Hamiltonian in the three relative coordinates. Since
$H_0$ and $W$ are both independent of $\alpha$, and $H(\alpha)$ is exactly linear in
it, each sector need only be constructed once and may then be swept over $\alpha$ at
the cost of a single eigensolve per value.

The basis is truncated by total oscillator quanta, retaining every state for which
$\sum_{k} (n_{a,k} + n_{b,k}) \leq q_\text{max}$. Because $H(\alpha)$ is non-Hermitian we cannot use a Lanczos method. We use a generalized Davidson iteration preconditioned by the diagonal of
$H_0$. 

Convergence in $q_\text{max}$ is not captured by a simple exponential or power law, and, the
transformed Hamiltonian being non-Hermitian, the truncated energies are not
variational. An extrapolation is therefore required, and we describe here the
procedure used to produce the $q_\text{max} \to \infty$ values of Table~\ref{table:energies}.

For $\alpha \lesssim 0.55$ the ground state is the linear branch~\cite{wu1984} at $E = 4 + 6\alpha$, and at $\alpha = 1$ it is the four-fermion
level $E = 8$. In both cases the exact eigenstate lies within the truncated basis
and the energy is obtained to machine precision at every $q_\text{max}$, so no
extrapolation is performed. For the remaining values of $\alpha$ the ground state
lies on a different branch, and we compute a series at
$q_\text{max} = 24, 28, \ldots, 56$ and fit it to

\begin{equation}
  \begin{aligned}
    E(q_\text{max}) = E_0 + c_1 q_\text{max}^{-c_2}
                          + c_3 q_\text{max}^{-c_4}\,,
    \label{eq:twoTermFit}
  \end{aligned}
\end{equation}

with all five parameters free and the amplitudes $c_1$ and $c_3$ unconstrained in
sign. The errors on the extrapolated energy are estimated via computing the spread of $E_0$'s obtained by fitting with one value of $q_\text{max}$ omitted and quoting the standard deviation. The multi-exponent fit is necessary due to the non-monotonic behavior of the energy as a function of $q_\text{max}$ for $\alpha=0.6$. This complicated non-monotonic behavior exemplifies the difficulties in studying continuum anyonic systems with ED; although our extrapolation gives energies extremely close to our VMC results, one cannot completely rule out the possibility that other non-monotonic behavior occurs at larger $q_\text{max}$ which would substantially change the prediction.

\begin{figure*}[!h]
    \centering
    \includegraphics[width=\textwidth]{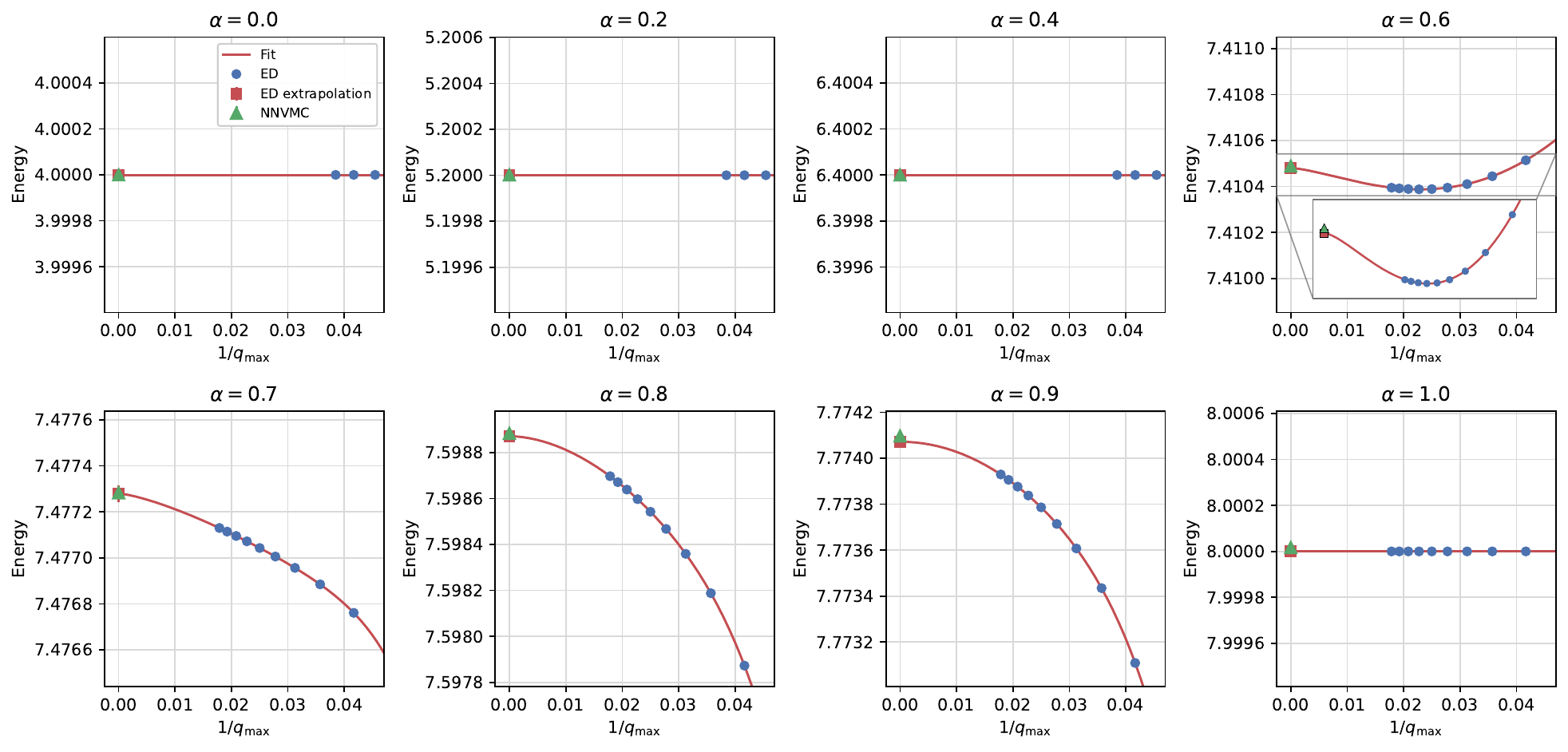}
    \caption{Exact diagonalization energy for $N = 4$ as a function of the maximum number of oscillator quanta above the ground state ($q_\text{max}$) retained in the calculation. We fit the data to $E_{\text{ED}}(q_\text{max}) = E_{0} + c_{1} \times q_\text{max}^{-c_{2}} + c_{3} \times q_\text{max}^{-c_{4}}$; a single power law cannot reproduce the non-monotonic approach seen at $\alpha = 0.6$. At $\alpha \in {0, 0.2, 0.4, 1}$ the ground state lies on the linear branch or at the free-fermion point and is obtained exactly at every $q_\text{max}$, while at $\alpha \in {0.7, 0.8, 0.9}$ the energy is still drifting in the fourth decimal place at the largest $q_\text{max}$ available. The extrapolation nonetheless determines $E_{0}$ to better than $4 \times 10^{-5}$, limited by the choice of fit rather than by the scatter of the data. The inset for $\alpha=0.6$ shows a zoomed-in version of the same data in order to more clearly demonstrate the non-monotonicity of the energy.}
    \label{fig:ED_extrapolation}
\end{figure*}

\section{Results for $N = 6$ anyons}
\label{chap:results_n6}
\begin{figure*}[!h]
    \centering
    \includegraphics[width=\textwidth]{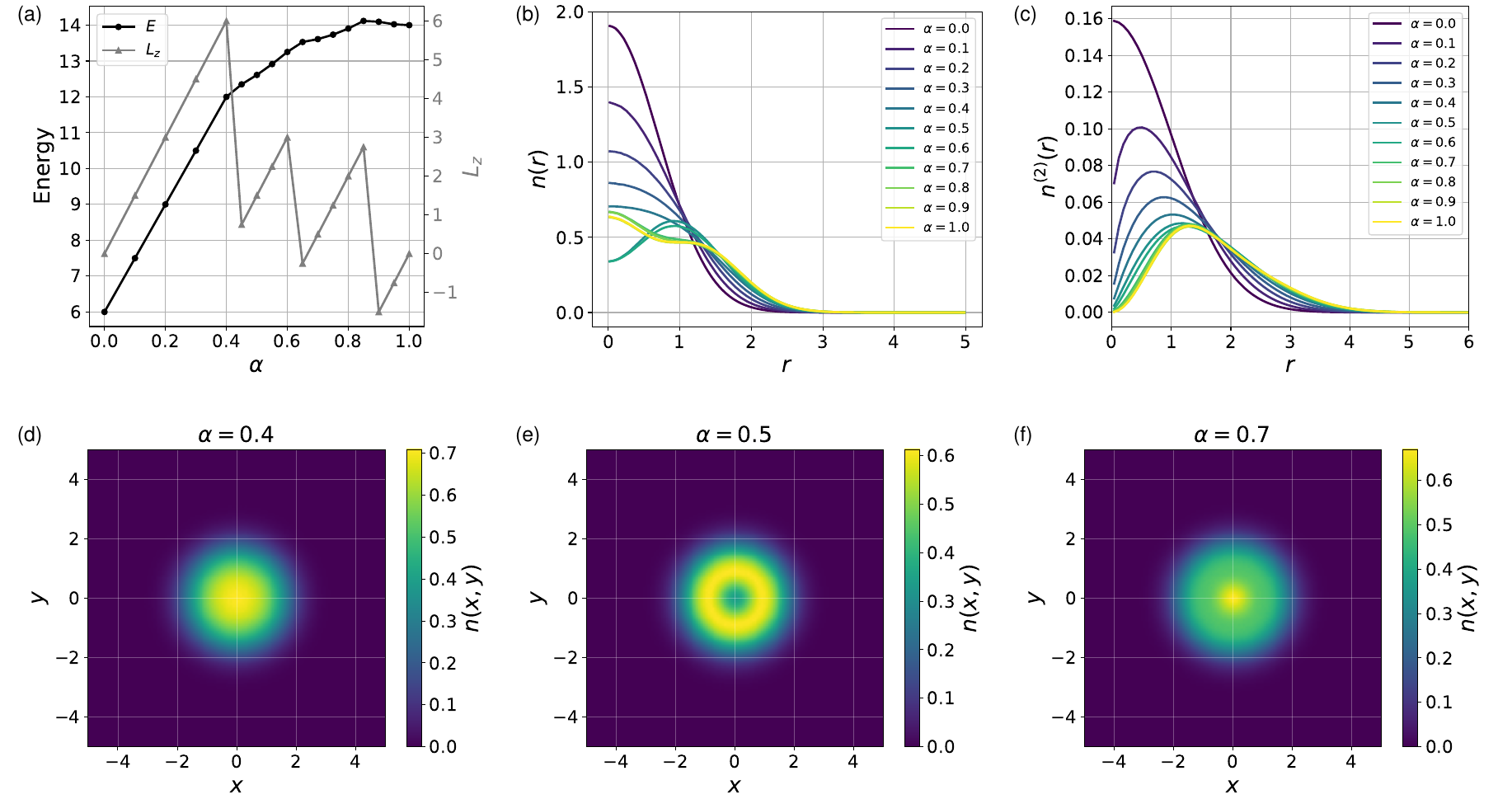}
    \caption{Results for $N = 6$ anyons. 
    Panel $(a)$ shows the ground state energy and the total angular momentum as a function of the exchange parameter $\alpha$.
    Error bars are not plotted as they are too small to be seen.
    $(b)$: Radial density $n(r)$ of the ground state wavefunction. At $\alpha \approx 0.5$, the density develops a hole in the center of the harmonic well. 
    $(c)$: Translationally averaged pair-density function $n^{(2)}(r)$ for a range of $\alpha$. 
    Panels $(d)$, $(e)$, and $(f)$ correspond to the real space density at $\alpha = 0.4$, $0.5$, and $0.7$, respectively. }
    \label{fig:n6}
\end{figure*}
In the main text, we present our results for $N = 10$, the largest system for which we attempted to find the spectrum as a function of $\alpha$.
In Fig.~\ref{fig:n6}, we present the results for $N = 6$, in which there are fewer level crossings, and the results look less cluttered.

\newpage
\section{Perturbation theory near the fermionic end}\label{chap: perturbation appendix}
Our goal here is to evaluate $\mathcal{M}$ in Eq.~\eqref{eq: M_PQ definition} analytically, so that the computation of the first-order energy correction coefficient $\Lambda_N$ becomes an efficient task of computing eigenvalues of several small matrices.

\subsubsection{Fermionic ground-state manifold}
To facilitate the calculation of many-body matrix element $\mathcal{M}_{PQ}$, let us adopt the following single-particle angular-momentum eigen-basis ($z=x+iy$, $\bar{z}=x-iy$):
\begin{equation}
    \varphi_{\bp}(z,\bar{z})
 =\frac{e^{-\abs{z}^2/2}}{\sqrt{\pi p_+!p_-!}}h_{p_+p_-}(z,\bar z),
\end{equation}
where each orbital is labeled by $\bp=(p_+,p_-) \in\mathbb{N}_0^2$ ($\mathbb{N}_0$ denotes the non-negative integers), and $h_{ab}(z,\bar z) = (-1)^{a+b} e^{\abs{z}^2} \partial^a_{\bar{z}}\partial^b_{z}e^{-\abs{z}^2}$ is known as the complex Hermite polynomial. The above orbitals constitute a complete orthonormal basis such that $\varphi_\bp$ has energy $\varepsilon_\bp = p_++p_-+1$ (in unit of $\hbar\omega$) and angular momentum $m_\bp = p_+-p_-$ (in unit of $\hbar$). 

\begin{figure}[t!]
    \centering
    \includegraphics[width=0.4\columnwidth]{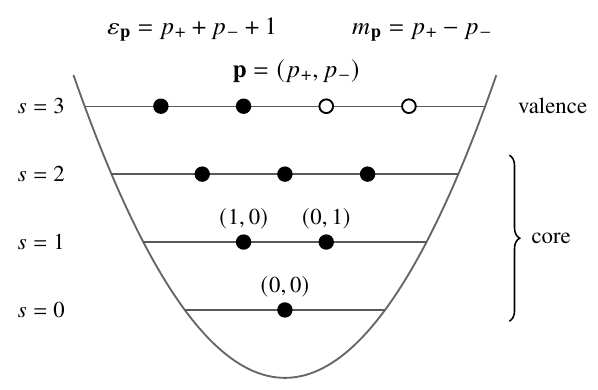}
    \caption{Schematic for the many-fermion ground state in a harmonic trap. The first-order correction to the many-anyon ground state energy, near the fermionic limit, is obtained from evaluating the matrix element in Eq.~\eqref{eq: M_PQ definition}}
    \label{fig: fermion harmonic trap setup}
\end{figure}

The $N$-fermion ground states are Slater-determinant states,
\begin{equation}\label{eq: N-fermion gs in harmonic trap}
    \Psi_P(z_1,\ldots,z_N)
 =\frac1{\sqrt{N!}}\det_{1\le a,b\le N}[\varphi_{\bp_b}(z_a)],
\end{equation}
where here and below we suppress the $\bar{z}$ argument for simplicity. Each many-body state is specified by an ordered list of occupied orbitals: $P=(\bp_1,\ldots,\bp_N)$. The ordering is stipulated to be increasing in $\varepsilon_\bp$, and for each fixed $\varepsilon_\bp$ arranged with a decreasing $m_\bp$. It is easy to see that the fermionic ground state energy and degeneracy, respectively, are
 \begin{equation}
E_N(1)=\frac{s(s+1)(2s+1)}6+r(s+1),\;\;
 D_N=\binom{s+1}{r},
 \label{eq:EF-D}
\end{equation}
where the integer $s \geq 0$ labels the highest occupied shell (with energy and degeneracy $s+1$) which satisfies $s(s+1)/2<N\le (s+1)(s+2)/2$, and $r=N-s(s+1)/2$ is the number of fermions filling the highest shell $s$. For large $N$, it is clear that the fermionic ground-state energy scales as 
\begin{equation}\label{eq: fermionic gs E scaling}
    E_N(1) \sim s^3/3\sim N^{3/2}\sqrt{8}/3.
\end{equation}
Furthermore, it is useful to separate the occupied orbitals into two categories: the ``core" orbitals below shell $s$ and the ``valence" orbitals in shell $s$. Figure~\ref{fig: fermion harmonic trap setup} shows an illustration for organizing the fermionic ground states. 

\subsubsection{Many-body matrix elements}
Combining Eq.~\eqref{eq: M_PQ definition} with Eq.~\eqref{eq: N-fermion gs in harmonic trap}, we obtain a compact form for the many-body matrix elements in terms of two-body matrix elements:
\begin{equation}\label{eq: M in terms of W}
    \mathcal{M}_{PQ} = \sum_{\substack{i<j,\ k<l\\
                 P_{ij}=Q_{kl}}}
 (-1)^{i+j+k+l}
 W_{\bp_i\bp_j,\bq_k\bq_l},
\end{equation}
where $P_{ij}$ corresponds to the list of occupied orbitals with $\bp_i$ and $\bp_j$ removed, and 
\begin{subequations}
\begin{align}
    W_{\bp\bq,\br\bs} &= \bra{\sigma_{\bp\bq}}\hat{V}_{12}\ket{\sigma_{\br\bs}},\\
    \sigma_{\bp\bq}(z_1,z_2)&=\frac{1}{\sqrt{2}}[\varphi_{\bp}(z_1)\varphi_{\bq}(z_2)
        -\varphi_{\bq}(z_1)\varphi_{\bp}(z_2)].
\end{align}
\end{subequations}
Energy and angular momentum constraints dictate that the two-body matrix elements $W_{\bp\bq,\br\bs}$ of concern must obey $p_++q_+ = r_++s_+$ and $p_-+q_- = r_-+s_-$. The two-body integral in $W_{\bp\bq,\br\bs}$ can be carried out analytically, resulting in 
\begin{equation}\label{eq: explicit form of W}
    W_{\bp\bq,\br\bs}
 =2\sum_{\substack{0\le i\le p,\ 0\le j\le q\\i+j\ \mathrm{odd}}}
 \sgn(i-j)\,C_{ij}^{\bp\bq}C_{ij}^{\br\bs}, 
\end{equation}
where we have defined $p\equiv p_++q_+$, $q\equiv p_-+q_-$, and 
\begin{subequations}\label{eq: explicit form of C}
    \begin{align}
        C_{ij}^{\bp\bq}
 &=
 \sqrt{\frac{(p-i)!(q-j)!i!j!}{2^{p+q}\;p_+!p_-!q_+!q_-!}}
 B_i^{p_+,q_+}B_j^{p_-,q_-},\\
  B_i^{a,b}&=\sum_{k=0}^{i}(-1)^{i-k}\binom ak\binom b{i-k}.
    \end{align}
\end{subequations}
To shed some light on Eq.~\eqref{eq: explicit form of W}, let us mention that $C_{ij}^{\bp\bq}$ is a combinatorial factor serving as an expansion coefficient for the two-fermion state $\sigma_{\bp\bq}(z_1,z_2)$ in terms of the center-of-mass $Z = (z_1+z_2)/\sqrt{2}$ and relative $u=(z_1-z_2)/\sqrt{2}$ coordinates:
\begin{equation}
    \sigma_{\bp\bq}
 =\sqrt2\sum_{\substack{0\le i\le p,\ 0\le j\le q\\i+j\ \mathrm{odd}}}
 C_{ij}^{\bp\bq}\,
 \varphi_{p-i,q-j}(Z)\varphi_{ij}(u).
\end{equation}
Only odd $i+j$ contribute due to the fermionic antisymmetry and the fact that $\varphi_{ij}(-u)= (-1)^{i+j}\varphi_{ij}(u)$, while the factor $\sgn(i-j)$ in Eq.~\eqref{eq: explicit form of W} is the result of the relative-coordinate integral. 

\begin{table}[h]
\caption{Two-body matrix element $ W_{\bp_i\bp_4, \bp_i\bp_4}$ for different combinations of core orbitals $\bp_{i=1,2,3}$ and valence orbital $\bp_4$.}
\begin{center}\label{table: W for N=4}
\begin{tabular}{c c c c}
\toprule
$\bp_{i=1,2,3}\; \backslash \;\bp_4$\;\;\;\; & \;$(2,0)$\; & \;$(1,1)$ \;&\; $(0,2)$\;\\
\midrule
$(0,0)$ & $1$ & $0$ & $-1$\\
$(1,0)$ & $1$ & $0$ & $-1/2$\\
$(0,1)$ & $1/2$ & $0$ & $-1$\\
\midrule
$\sum_i W_{\bp_i\bp_4, \bp_i\bp_4}=$\;\;\; & $5/2$ & $0$ & $-5/2$\\
\bottomrule
\end{tabular}
\end{center}
\end{table}
\subsubsection{$\Lambda_N$ as a benchmark}
Combining Eqs.~\eqref{eq: Lambda_N from M},~\eqref{eq: M in terms of W},~\eqref{eq: explicit form of W} and ~\eqref{eq: explicit form of C}, we have all the ingredients needed to compute $\Lambda_N$. Notice that the summation in Eq.~\eqref{eq: M in terms of W} can be largely reduced by discarding terms where both $\bp_i$ and $\bp_j$ (accordingly, $\bq_k$ and $\bq_l$) are core orbitals, as the sum over all these terms must vanish by the fact that $\hat{V}_{12}$ is odd under time-reversal.

As an explicit example, take $N=4$, where we have three degenerate fermionic ground states that share the same core orbitals: $\bp_1=(0,0)$, $\bp_2=(1,0)$ and $\bp_3=(0,1)$, and distinguished by their single valence orbital: $\bp_4=(2,0),\;(1,1),\;(0,2)$, which give them distinct total angular momenta (and hence $\mathcal{M}$ is diagonal as constructed). The relevant terms in Eq.~\eqref{eq: M in terms of W} are $W_{\bp_i\bp_4, \bp_i\bp_4}$ ($i=1,2,3$), which are tabulated in Table~\ref{table: W for N=4}. From this we obtain $\Lambda_4=5/2$, consistent with the result obtained earlier using the non-relativistic conformal field theory technique \cite{nishida2007}.

With the method developed above, we can systematically obtain $\Lambda_N$ for a range of $N$. In Table~\ref{tab:lambda-nn-vmc}, for a selected set of $N$ we have documented a comparison between the exact perturbation theory results with our quadratic fitting results obtained from the anyonic NNVMC simulations with $\alpha=0.98, 0.99, 0.995, 0.9975$ and $1$. The observed close numerical match, which is also shown in Fig.~\ref{fig: lambda_N}, demonstrates the effectiveness of our AnyNet ansatze for simulating intermediate scale anyonic systems. We further observe from Fig.~\ref{fig: lambda_N} that $\Lambda_N$ scales as $N^{3/2}$ at large $N$, similar to the fermionic ground state energy $E_N(1)$ in Eq.~\eqref{eq: fermionic gs E scaling}. This is a direct demonstration of an effective attraction (or a weakened Pauli exclusion) in an almost-fermionic anyon gas. 

\section{Additional data}
\label{chap:additional_data}
In this section, we present some of the numerical values for the energy and angular momentum of our variational wavefunctions for each system.
For a full set of energies and angular momenta, see the supplementary information.
\begin{table}[htbp]
  \centering
  \footnotesize
  \setlength{\tabcolsep}{2.5pt}
  \sisetup{separate-uncertainty=false, group-digits=none}
  \caption{Variational (a) energies $E$ and (b) angular momenta $\langle L_z\rangle$ for $N$ anyons in a harmonic trap, as a function of the statistical parameter $\alpha$ (units $\hbar=m=\omega=1$).}
  \label{tab:energies}
  \textbf{(a) Energy $E$}\par\smallskip
  \begin{tabular}{c S[table-format=1.6(1)] S[table-format=1.7(1)] S[table-format=2.7(1)] S[table-format=2.7(1)] S[table-format=2.7(1)] S[table-format=2.6(1)]}
    \toprule
    $\alpha$ & {$N=3$} & {$N=4$} & {$N=5$} & {$N=6$} & {$N=8$} & {$N=10$} \\
    \midrule
    0.0 & 2.999997(2) & 4.0000004(4) & 5.000006(6) & 6.0000001(6) & 8.000003(3) & 9.99995(5) \\
    0.1 & 3.300009(5) & 4.599999(1) & 6.000001(1) & 7.5000001(8) & 10.8000010(3) & 14.500001(5) \\
    0.2 & 3.600014(7) & 5.200002(2) & 7.0000015(4) & 9.0000016(8) & 13.6000006(5) & 18.99999(1) \\
    0.3 & 3.900000(6) & 5.800002(2) & 7.999998(2) & 10.500002(2) & 16.4000018(3) & 22.6512(1) \\
    0.4 & 4.200013(7) & 6.400001(4) & 9.000006(4) & 12.000002(2) & 19.1999991(5) & 25.14487(6) \\
    0.5 & 4.500006(7) & 7.000006(4) & 9.81754(1) & 12.61134(2) & 19.46797(3) & 27.01276(5) \\
    0.6 & 4.800022(5) & 7.410489(7) & 10.079177(9) & 13.25285(1) & 20.24540(3) & 28.11512(3) \\
    0.7 & 5.10003(1) & 7.477283(5) & 10.436656(7) & 13.61027(1) & 20.85210(4) & 29.13769(3) \\
    0.8 & 5.051217(6) & 7.598884(4) & 10.660542(6) & 13.904836(7) & 21.46856(1) & 29.82747(2) \\
    0.9 & 5.012927(5) & 7.774097(4) & 10.776775(4) & 14.094794(3) & 21.532750(5) & 30.10778(2) \\
    1.0 & 5.000012(5) & 8.000018(4) & 11.000025(3) & 14.000004(2) & 22.000000(1) & 30.000061(9) \\
    \bottomrule
  \end{tabular}

  \bigskip

  \textbf{(b) Angular momentum $\langle L_z\rangle$}\par\smallskip
  \begin{tabular}{c S[table-format=-1.7(1)] S[table-format=-1.10(1)] S[table-format=-1.8(1)]}
    \toprule
    $\alpha$ & {$N=3$} & {$N=6$} & {$N=10$} \\
    \midrule
    0.0 & 0.0000000(1) & -0.0000001333(5) & 0.00000000(4) \\
    0.1 & 0.2999998(2) & 1.49999996(1) & 4.50000006(4) \\
    0.2 & 0.5999995(3) & 3.00000005(4) & 9.00000010(5) \\
    0.3 & 0.8999995(3) & 4.49999999(3) & 3.499868(6) \\
    0.4 & 1.2000008(5) & 5.99999987(8) & 7.999872(6) \\
    0.5 & 1.4999996(3) & 1.499993(1) & 4.499893(6) \\
    0.6 & 1.7999989(4) & 2.999973(2) & 2.999930(4) \\
    0.7 & 2.0999934(6) & 0.499976(2) & 3.499881(5) \\
    0.8 & -0.6000000(3) & 1.999986(1) & 0.999985(3) \\
    0.9 & -0.2999996(2) & -1.4999977(6) & 2.499982(3) \\
    1.0 & 0.0000001(2) & 0.0000001(2) & -0.0000006(6) \\
    \bottomrule
  \end{tabular}
\end{table}

\begin{table}[htbp]
  \centering
  \footnotesize
  \setlength{\tabcolsep}{2.4pt}
  \sisetup{separate-uncertainty=false, group-digits=none}
  \caption{Variational energies $E$ and angular momenta $\langle L_z\rangle$ for $N$ anyons in a harmonic trap, for statistical parameters $\alpha=1/2$, $2/3$ and $3/4$ (units $\hbar=m=\omega=1$). The number in parentheses is the statistical uncertainty in the last digit.
  We also note that for $N = 7$, there is a level crossing very near $\alpha = 3/4$, which makes it hard to converge the angular momentum eigenvalue.}
  \label{tab:energies_frac}
  \begin{tabular}{c S[table-format=2.6(1)] S[table-format=-1.7(1)] S[table-format=2.6(1)] S[table-format=-1.7(1)] S[table-format=2.6(1)] S[table-format=-1.7(1)]}
    \toprule
    & \multicolumn{2}{c}{$\alpha=1/2$} & \multicolumn{2}{c}{$\alpha=2/3$} & \multicolumn{2}{c}{$\alpha=3/4$} \\
    \cmidrule(lr){2-3} \cmidrule(lr){4-5} \cmidrule(lr){6-7}
    $N$ & {$E$} & {$\langle L_z\rangle$} & {$E$} & {$\langle L_z\rangle$} & {$E$} & {$\langle L_z\rangle$} \\
    \midrule
    2 & 2.500002(4) & 0.5000000(1) & 2.666667(2) & 0.6666666(1) & 2.750004(3) & 0.7499999(1) \\
    3 & 4.500015(6) & 1.4999986(4) & 5.000011(8) & 1.9999969(6) & 5.079515(8) & -0.7499977(5) \\
    4 & 7.000006(4) & 3.000004(1) & 7.44901(2) & 0.000000(1) & 7.53134(1) & 0.4999993(7) \\
    5 & 9.81762(2) & -0.000003(1) & 10.30740(2) & 1.666660(1) & 10.64346(1) & -0.499827(4) \\
    6 & 12.61143(2) & 1.499994(2) & 13.55210(2) & -0.000004(1) & 13.73514(1) & 1.249995(1) \\
    7 & 15.81918(2) & 3.499986(2) & 16.93829(1) & 1.999992(1) & 17.46589(1) & 1.5411(2) \\
    8 & 19.46797(3) & 5.999959(3) & 20.70225(2) & 0.666660(2) & 21.15270(2) & 2.999963(3) \\
    9 & 23.12718(3) & 1.999955(3) & 24.57243(2) & 2.999968(3) & 25.25762(2) & -1.000021(3) \\
    10 & 27.01276(5) & 4.499893(6) & 28.83805(3) & 1.999932(4) & 29.49371(2) & 1.749961(4) \\
    11 & 31.25165(6) & 0.499808(9) & 33.16552(3) & 4.666573(6) & 33.92712(3) & 1.249957(5) \\
    12 & 35.38585(6) & 2.999759(9) & 37.89081(4) & 3.999716(9) & 38.70499(4) & 4.49966(1) \\
    13 & 39.8924(1) & 5.99863(3) & 42.66048(4) & 6.99968(1) & 43.59587(4) & 0.499843(8) \\
    14 & 44.6073(1) & 1.49803(3) & 47.76995(7) & 0.66614(2) & 48.58440(6) & 0.24951(1) \\
    15 & 49.3247(1) & 4.49841(3) & 52.91898(8) & 3.99934(2) & 53.82549(6) & 3.74967(1) \\
    16 & 54.4190(2) & 7.99614(4) & 58.23422(9) & 2.99890(2) & 59.31970(7) & 3.99947(1) \\
    \bottomrule
  \end{tabular}
\end{table}

\begin{table}[htbp]
  \centering
  \footnotesize
  \setlength{\tabcolsep}{3pt}
  \sisetup{separate-uncertainty=false, group-digits=none}
  \caption{Variational energies $E$ and angular momenta $\langle L_z\rangle$ for $N$ anyons in a harmonic trap, for statistical parameters $\alpha$ close to the fermionic limit $\alpha=1$ (units $\hbar=m=\omega=1$). The number in parentheses is the statistical uncertainty in the last digit; a dash marks $\langle L_z\rangle$ at $\alpha=1$, which is not well defined since the open-shell fermionic ground state is degenerate.}
  \label{tab:energies_near_fermion}
  \begin{tabular}{S[table-format=1.4] S[table-format=2.6(1)] S[table-format=1.7(1)] S[table-format=2.5(1)] S[table-format=1.6(1)] S[table-format=2.6(1)] S[table-format=1.6(1)]}
    \toprule
    & \multicolumn{2}{c}{$N=5$} & \multicolumn{2}{c}{$N=8$} & \multicolumn{2}{c}{$N=12$} \\
    \cmidrule(lr){2-3} \cmidrule(lr){4-5} \cmidrule(lr){6-7}
    {$\alpha$} & {$E$} & {$\langle L_z\rangle$} & {$E$} & {$\langle L_z\rangle$} & {$E$} & {$\langle L_z\rangle$} \\
    \midrule
    0.98 & 10.947067(8) & 1.7999968(8) & 21.87535(1) & 3.439967(3) & 39.787773(7) & 4.679966(4) \\
    0.99 & 10.973035(7) & 1.899987(1) & 21.93586(1) & 3.719851(4) & 39.888117(5) & 5.339946(4) \\
    0.995 & 10.986406(7) & 1.949963(2) & 21.96754(1) & 3.859836(5) & 39.942628(4) & 5.669953(4) \\
    0.9975 & 10.993183(8) & 1.974705(5) & 21.98375(1) & 3.92948(1) & 39.970972(4) & 5.834817(7) \\
    1 & 11.000023(8) & {--} & 22.00018(2) & {--} & 40.000028(5) & {--} \\
    \bottomrule
  \end{tabular}

  \bigskip

  \begin{tabular}{S[table-format=1.4] S[table-format=2.5(1)] S[table-format=1.6(1)] S[table-format=2.5(1)] S[table-format=1.6(1)]}
    \toprule
    & \multicolumn{2}{c}{$N=16$} & \multicolumn{2}{c}{$N=20$} \\
    \cmidrule(lr){2-3} \cmidrule(lr){4-5}
    {$\alpha$} & {$E$} & {$\langle L_z\rangle$} & {$E$} & {$\langle L_z\rangle$} \\
    \midrule
    0.98 & 60.83105(3) & 2.599961(6) & 84.84061(4) & 1.199936(8) \\
    0.99 & 60.90329(3) & 3.799919(9) & 84.89824(5) & 3.099954(7) \\
    0.995 & 60.94867(3) & 4.399897(8) & 84.94407(4) & 4.049932(7) \\
    0.9975 & 60.97385(3) & 4.69946(1) & 84.97100(4) & 4.52464(2) \\
    1 & 61.00056(3) & {--} & 85.00120(4) & {--} \\
    \bottomrule
  \end{tabular}
\end{table}

\begin{table}[htbp]
  \centering
  \footnotesize
  \setlength{\tabcolsep}{3pt}
  \sisetup{separate-uncertainty=false, group-digits=none}
  \caption{Variational energies $E$, angular momenta $\langle L_z\rangle$ and mean-square radii $\langle r^2\rangle$ of anyonic atoms with $N$ anyons and nuclear charge $Z$, as a function of the statistical parameter $\alpha$. $N=Z$ is the neutral atom and $N=Z+1$ the atom with one extra anyon. Panels (a) and (b) include the anyon--anyon Coulomb interaction; panel (c) shows the neutral atoms without it. The number in parentheses is the statistical uncertainty in the last digit.
  Note that observables for the non-neutral atoms where the additional anyon does not bind are not meaningful, as those states cannot be converged with an exponentially decaying ansatz.
  Angular momentum eigenvalues for open shells are similarly not meaningful and are not shown.}
  \label{tab:anyonic_atoms}
  \textbf{(a) Interacting, $Z=2$}\par\smallskip
  \begin{tabular}{S[table-format=1.1] S[table-format=-2.6(1)] S[table-format=-1.8(1)] S[table-format=1.4(1)] S[table-format=-2.5(1)] S[table-format=-1.7(1)] S[table-format=2.4(1)]}
    \toprule
    & \multicolumn{3}{c}{$N=2$, $Z=2$} & \multicolumn{3}{c}{$N=3$, $Z=2$} \\
    \cmidrule(lr){2-4} \cmidrule(lr){5-7}
    {$\alpha$} & {$E$} & {$\langle L_z\rangle$} & {$\langle r^2\rangle$} & {$E$} & {$\langle L_z\rangle$} & {$\langle r^2\rangle$} \\
    \midrule
    0.0 & -11.89982(3) & 0.00000009(6) & 0.2992(1) & -12.9360(2) & -0.0000001(2) & 0.9111(5) \\
    0.1 & -11.14723(8) & 0.10000006(4) & 0.3716(2) & -11.6409(1) & 0.2999996(8) & 1.576(1) \\
    0.2 & -10.45888(6) & 0.19999995(4) & 0.4729(2) & -10.6109(1) & 0.599991(2) & 3.447(5) \\
    0.3 & -9.8598(1) & 0.2999998(1) & 0.6201(3) & -9.8574(1) & 0.90010(1) & 11.26(2) \\
    0.4 & -9.36033(6) & 0.39999989(8) & 0.8482(5) & -9.3293(1) & 1.16753(2) & 36.72(5) \\
    0.5 & -8.96432(6) & 0.5000000(1) & 1.2339(8) & -8.9333(1) & 0.5519(6) & 54.28(5) \\
    0.6 & -8.67091(5) & 0.5999999(2) & 1.928(2) & -8.6396(1) & -0.1692(3) & 54.66(7) \\
    0.7 & -8.47165(3) & 0.6999996(3) & 3.178(3) & -8.45617(7) & 0.100029(6) & 28.51(3) \\
    0.8 & -8.34740(3) & 0.7999995(4) & 5.173(4) & -8.37895(5) & 0.399986(5) & 24.76(2) \\
    0.9 & -8.29519(1) & -0.099999(2) & 8.715(9) & -8.33955(4) & 0.699992(6) & 27.57(3) \\
    1.0 & -8.295961(5) & 0.0000002(2) & 8.668(7) & -8.31075(6) & {--} & 30.83(4) \\
    \bottomrule
  \end{tabular}

  \bigskip

  \textbf{(b) Interacting, $Z=3$}\par\smallskip
  \begin{tabular}{S[table-format=1.1] S[table-format=-2.5(1)] S[table-format=-1.8(1)] S[table-format=2.4(1)] S[table-format=-2.4(1)] S[table-format=-1.7(1)] S[table-format=2.4(1)]}
    \toprule
    & \multicolumn{3}{c}{$N=3$, $Z=3$} & \multicolumn{3}{c}{$N=4$, $Z=3$} \\
    \cmidrule(lr){2-4} \cmidrule(lr){5-7}
    {$\alpha$} & {$E$} & {$\langle L_z\rangle$} & {$\langle r^2\rangle$} & {$E$} & {$\langle L_z\rangle$} & {$\langle r^2\rangle$} \\
    \midrule
    0.0 & -35.8823(1) & 0.00000007(8) & 0.2421(1) & -38.2653(1) & -0.0000001(2) & 0.5325(4) \\
    0.1 & -31.3076(3) & 0.2999998(2) & 0.3834(2) & -32.0629(2) & 0.59999(1) & 1.246(1) \\
    0.2 & -27.7485(3) & 0.5999990(5) & 0.6358(4) & -27.8364(2) & 1.19987(2) & 4.56(1) \\
    0.3 & -25.0670(2) & 0.8999989(5) & 1.1442(8) & -25.0223(3) & 1.79921(7) & 21.68(3) \\
    0.4 & -23.1023(4) & 1.199980(3) & 2.283(3) & -23.0549(3) & 0.4222(5) & 43.71(5) \\
    0.5 & -21.6916(3) & 1.499972(7) & 4.913(5) & -21.6707(4) & 1.00117(6) & 31.90(3) \\
    0.6 & -20.7065(2) & -0.19978(3) & 10.46(1) & -20.7192(2) & -0.39934(4) & 37.80(4) \\
    0.7 & -20.1726(1) & 0.100007(3) & 6.740(8) & -20.1506(2) & 0.20236(6) & 40.85(4) \\
    0.8 & -19.86736(8) & 0.400000(1) & 5.799(6) & -19.8268(2) & 0.79906(6) & 52.49(4) \\
    0.9 & -19.67541(5) & 0.6999993(7) & 6.221(5) & -19.6301(2) & -0.59930(9) & 27.52(4) \\
    1.0 & -19.52947(1) & {--} & 7.221(6) & -19.5870(1) & -0.000069(6) & 18.22(2) \\
    \bottomrule
  \end{tabular}

  \bigskip

  \textbf{(c) Non-interacting}\par\smallskip
  \begin{tabular}{S[table-format=1.1] S[table-format=-2.6(1)] S[table-format=-1.8(1)] S[table-format=-2.5(1)] S[table-format=1.8(1)]}
    \toprule
    & \multicolumn{2}{c}{$N=2$, $Z=2$} & \multicolumn{2}{c}{$N=3$, $Z=3$} \\
    \cmidrule(lr){2-3} \cmidrule(lr){4-5}
    {$\alpha$} & {$E$} & {$\langle L_z\rangle$} & {$E$} & {$\langle L_z\rangle$} \\
    \midrule
    0.0 & -16.00001(2) & -0.0000002(2) & -54.00001(5) & 0.00000001(6) \\
    0.1 & -14.5300(2) & 0.10000003(5) & -44.7843(3) & 0.299999(1) \\
    0.2 & -13.29332(8) & 0.19999992(7) & -38.0545(3) & 0.5999992(3) \\
    0.3 & -12.25582(7) & 0.29999994(5) & -33.0792(3) & 0.8999999(2) \\
    0.4 & -11.39147(7) & 0.39999994(5) & -29.3797(3) & 1.1999998(4) \\
    0.5 & -10.68044(6) & 0.49999996(6) & -26.6326(2) & 1.4999986(5) \\
    0.6 & -10.10692(5) & 0.59999988(9) & -24.6040(2) & 1.799992(1) \\
    0.7 & -9.65699(4) & 0.70000003(8) & -23.4198(2) & 0.0999994(8) \\
    0.8 & -9.31638(3) & 0.7999994(2) & -22.8142(1) & 0.3999995(3) \\
    0.9 & -9.06717(2) & 0.8999998(2) & -22.3647(1) & 0.6999996(4) \\
    1.0 & -8.888890(5) & {--} & -22.00003(3) & {--} \\
    \bottomrule
  \end{tabular}
\end{table}

\end{document}